\documentclass[11pt,a4paper]{article}
\pdfoutput=1

\usepackage{jheppub}
\usepackage{mathrsfs}

\def\S{{\mathbb S}}

\newcommand{\alg}[1]{\mathfrak{#1}}

\def\ads{{\rm AdS}_5\times {\rm S}^5}

\newcommand{\tl}[1]{\tilde{#1}}

\def\ads{{\rm AdS}_5\times {\rm S}^5}

\title{The spectral problem for strings on twisted $\mbox{AdS}_5 \times \mbox{S}^5$}

\author[a]{Marius de Leeuw}
\author[b]{and Stijn J. van Tongeren}
\affiliation[a]{ETH Z\"urich, Institut f\"ur Theoretische Physik, \\Wolfgang-Pauli-Str.\ 27, CH-8093 Zurich, Switzerland}
\affiliation[b]{Institute for Theoretical
Physics and Spinoza Institute,\\Utrecht University, 3508 TD
Utrecht, The Netherlands}

\emailAdd{deleeuwm@phys.ethz.ch}
\emailAdd{s.j.vantongeren@uu.nl}

\abstract{We discuss the spectral problem for integrable superstrings on generically twisted $\ads$, meaning all its orbifolds and TsT transformed versions. We explicitly give the asymptotic description of these theories through a twisted transfer matrix, and carefully match the geometric deformations with twists allowed by integrability. We then discuss the mirror TBA equations that describe these theories at finite size. This unifies the treatment of various specific deformations previously considered in this setting, and extends it to completely general twists.}

\begin{document}

\begin{flushright}\small{ITP-UU-12/01\\SPIN-12/01}\end{flushright}

\maketitle

\section{Introduction}

This paper provides a unified description of the presently known integrable deformations of the $\ads$ superstring, from their asymptotic spectra through the Bethe ansatz equations and the twisted transfer matrix to their finite size description in terms of the mirror thermodynamic Bethe ansatz. Through the AdS/CFT correspondence \cite{Maldacena:1997re} these theories provide a description of the correspondingly deformed $\mathcal{N}=4$ supersymmetric Yang-Mills theories at finite 't Hooft coupling.

The integrability of the free superstring on $\ads$ and correspondingly planar $\mathcal{N}=4$ supersymmetric Yang-Mills theory (SYM) are by now well understood topics \cite{Arutyunov:2009ga,Beisert:2010jr}. From the point of view of AdS/CFT these theories can also be deformed, thereby extending this canonical example of AdS/CFT to less symmetric theories in a very controlled manner. The dualities fall into two classes: strings on orbifolds of $\ads$ dual to `orbifolds'\footnote{Orbifolding the sphere results in a quiver structure in the dual gauge theory.} of $\mathcal{N}=4$ SYM, as originally proposed in \cite{Kachru:1998ys}, and marginal deformations of $\mathcal{N}=4$ SYM dual to strings on Lunin-Maldacena-type backgrounds \cite{hep-th/0502086}. Due to the controlled nature of these deformations, the tools of integrability in the original theories can be adapted to these theories as well\footnote{There are still open questions regarding the finite size integrability of $\gamma$-deformed superstrings. In \cite{Frolov:2005ty,Frolov:2005iq} a class of finite-gap solutions were found, but it is not clear how to extend this to incorporate all finite-gap solutions for these deformations.}, as recently reviewed in \cite{Zoubos:2010kh}.

In this paper we focus mostly on aspects of the finite size integrability of these theories, following the more recent developments of finite size integrability of the undeformed superstring. Via the mirror trick \cite{ITEP-89-144} the spectrum of the superstring at finite size is described as the spectrum of its mirror theory in infinite size but at a finite temperature, which is a problem that can be tackled via the thermodynamic Bethe ansatz (TBA). The derivation of the mirror TBA equations for the $\ads$ superstring\footnote{For recent reviews of the TBA techniques in this context see for example \cite{Kuniba2,Bajnok:2010ke}.} is based on the corresponding string hypothesis \cite{Takahashi72} formulated in \cite{Arutyunov:2009zu} following from the corresponding mirror Bethe-Yang (BY) equations \cite{Arutyunov:2007tc}. This led to the formulation of the ground state \cite{AF09b,BFT,GKKV09} and excited state TBA equations for string states with both real \cite{GKKV09,AFS09,BH10b,Sfondrini:2011rr} and more recently complex momenta \cite{Arutyunov:2011mk}. As a particularly nice result, the TBA equations for the Konishi operator were shown to agree numerically \cite{AFS10} and analytically \cite{BH10a,BH10b} with L\"uscher's perturbative treatment \cite{BJ08,BJ09,LRV09,Janik:2010kd} and at four loops with explicit field-theoretic computations \cite{Sieg,Velizhanin:2008jd}. Let us also note that there have been important recent developments
\cite{Suzuki:2011dj,Gromov:2011cx} towards a description of the spectral problem through a finite set of non-linear integral equations (NLIE) complementing the TBA approach.

Coming back to deformed theories, in this paper we would like to apply the same type of ideas to them. As we will see below, this turns out to be possible because both types of deformations can be described through the original model with quasi-periodic boundary conditions. The fact that this is possible for strings on Lunin-Maldacena-type backgrounds with real deformation parameters was shown in \cite{hep-th/0503201} by linking these deformations to so-called TsT transformations, and was subsequently used there to find the Lax pair demonstrating the classical integrability of these deformed string theories. The effect of such TsT transformations for the full superstring coset model were discussed in \cite{hep-th/0512253} and for the Bethe ansatz equations of $\mathcal{N}=4$ SYM in \cite{Beisert:2005if}. For the case of orbifolds the description through the original model with quasi-periodic boundary conditions is simply the natural one, leaving nothing more to discuss, and gives their Bethe ansatz equations \cite{Beisert:2005he} similarly to the Lunin-Maldacena case.

Continuing in the direction of finite size integrability, the next step is to consider the transfer matrix. Integrable models with non-trivial boundary conditions can be described through a twisted transfer matrix \cite{E-11-86}, which was previously considered in the present context for deformations of the sphere \cite{arXiv:1009.4118}. At the same time it is possible to do a Drinfeld-Reshetikhin type twist on the $S$-matrix and derive a twisted transfer matrix and Bethe equations via this route\footnote{This concerns backgrounds obtained by TsT transformations; for orbifolds there is only the boundary condition point of view.}. This was considered in \cite{Ahn:2010yv,Ahn:2011} for deformations of the sphere and was found to be in full agreement with our approach. Moreover, it also appears to be possible to derive the twisted transfer matrix from a generating functional with a number of free twist parameters, by constraining the resulting transfer matrix to yield real Y-functions and reproduce the known twisted Bethe equations. This approach was successfully applied in \cite{Gromov:2010dy} to reproduce wrapping corrections in $\beta$-deformed SYM and it was subsequently shown that our twisted transfer matrix gives the same perturbative results \cite{arXiv:1009.4118}. In \cite{Beccaria:2011qd,Beccaria:2011qderr}\footnote{Note that unfortunately an outdated version of this paper was originally published; the second version on the arxiv and the recent erratum in JHEP \cite{Beccaria:2011qderr} correct this.} a variation of this approach was applied to abelian orbifolds of the sphere, in agreement with results of \cite{arXiv:1009.4118} where direct comparison is possible. The mirror TBA approach has also been successfully applied for specific deformations, namely orbifolds of the sphere in \cite{deLeeuw:2011rw}, and has in fact been carefully derived for $\gamma$-deformations of the sphere in \cite{arXiv:1108.4914}. Importantly, in \cite{arXiv:1108.4914} it was shown that the ground state energy in the TBA approach agrees with L\"uscher's approach at full double wrapping order.

In this paper we unify the treatment of these deformed theories in terms of one integrable model with varying boundary conditions. We will not restrict ourselves to deformations of the sphere but rather aim for full generality\footnote{Another interesting class of deformations of the underlying integrable structure is that of quantum deformations \cite{Beisert:2008tw,Beisert:2010kk,Beisert:2011wq,deLeeuw:2011jr}, but this takes the theory out of the realm of AdS/CFT.} and in fact (implicitly) treat all (sensible) TsT transformations and reasonable orbifolds of anti-de Sitter space. In other words we will treat all currently known integrable deformations simultaneously, including ones not concretely studied before.

As a first result we derive the generically twisted transfer matrix in full detail, and show that it immediately reproduces all known deformed asymptotic Bethe equations for specific deformations. For generality we also explicitly work out the generating functional with twist parameters as introduced in \cite{Gromov:2010dy} in an appendix and show that these parameters can be naturally identified with the twist parameters in our twisted transfer matrix. Doing so we show directly that our twisted transfer matrix agrees with the twisted generating functional of \cite{Gromov:2010dy} for deformations of the sphere, and extend this approach to deformations of anti-de Sitter space. Because of different conventions and corresponding choices of twists \cite{Beccaria:2011qd} does not directly line up with this approach; nonetheless explicit results agree. In order to derive the deformed mirror TBA equations \footnote{Once fully developed, the NLIE approach should be amenable to this type of deformations as well.} we realize that upon doing the mirror trick the non-trivial boundary conditions turn into a defect operator in the partition function of the theory \cite{ITEP-89-144} without modifying the string hypothesis or mirror BY equations. This means chemical potentials appear in the mirror TBA equations, but they are otherwise unchanged. As we will discuss, it is not hard to obtain the asymptotic behavior of the Y-functions from these equations, and this behavior naturally agrees with the asymptotic solution as constructed from the transfer matrices. In fact, as we show explicitly in an appendix, as expected the chemical potentials disappear from the simplified and hybrid forms of the TBA equations, meaning in particular that the resulting $Y$-system is unchanged for any of the deformations we consider. The twisted transfer matrix we find plays an important role also in the TBA setting, since it allows us to obtain excited state TBA equations by the contour deformation trick\footnote{This trick is inspired by the ideas of \cite{DT96,BLZe}.} \cite{AFS09,arXiv:1103.2708}.

Apart from providing a fully general and explicit set-up, we also briefly consider a few concrete applications. First of all we consider single wrapping corrections to the ground state energy given generic deformations, meaning we break supersymmetry completely as well as part of the conformal symmetry. The energy correction has a notably different functional form from previously studied cases owing to the breaking of conformal symmetry. The energy correction as well as the Y-functions also nicely show the partial restoration of supersymmetry in particular cases as we will discuss. Interestingly, by considering deformations in $\mbox{AdS}_5$ we find a regularization of the ground state energy of the original string theory which does not involve the curious divergence of the ground state energy at length $J=2$ previously observed in \cite{arXiv:0906.0499}. Moreover, we show that this divergence which was subsequently observed for non-supersymmetric orbifolds of the sphere \cite{deLeeuw:2011rw} as well as $\gamma$-deformations \cite{arXiv:1108.4914} is in fact a generic feature of the ground state energy at length two, appearing as soon as we attempt to take the $\mbox{AdS}_5$ deformations to zero. The origin of this divergence therefore remains an interesting open question. Furthermore, we explicitly indicate a class of theories with completely broken supersymmetry which nonetheless appear to have a `protected' ground state with zero energy.

We also consider wrapping corrections for two particle states in the $\alg{sl}(2)$ sector for a $\mathbb{Z}_4$ orbifold of $\mbox{AdS}_5$ that breaks all supersymmetry. These wrapping corrections quickly become rather unpleasant, and we only give their explicit form for the lightest two operators. This should perhaps not be too surprising since deformations of $\mbox{AdS}_5$ should have rather dramatic effects on the dual gauge theory.

This paper is organized as follows. We begin by discussing the types of integrable deformations of the $\ads$ superstring and how they can be described through the undeformed string theory with quasi-periodic boundary conditions. This is followed by a discussion of the implementation of these boundary conditions (hence the deformations) in the integrable structure of the theory via the twisted transfer matrix. Then in section \ref{sec:explicitmodels} we give concrete descriptions of specific deformations we can consider and make contact with existing literature where possible. Next, we show how these deformations can be incorporated naturally in the mirror TBA. This is followed by a discussion of the ground state in generically deformed theories. In section \ref{sec:adsorb} we give an exemplary discussion of wrapping corrections for a particular orbifold of $\mbox{AdS}_5$. Finally we briefly summarize our results. Some technical details and discussions have been relegated to an appendix.

\section{Integrable deformations of $\ads$}
\label{sec:introdef}

There are two classes of known integrable deformations of the $\ads$ superstring, and they are obtained by deforming the background space-time the superstring moves in. The first of these is a class of backgrounds obtained by orbifolding the space-time by a discrete subgroup of the isometry group of $\ads$. Secondly, we can also perform a sequence of so-called $T$-duality - shift - $T$-duality (TsT) transformations, giving a string theory on a TsT transformed background. The nice feature of these deformations is that both can be described in terms of the original string theory, where the deformation is accounted for by quasi-periodic boundary conditions for the string fields, as we will discuss below. This is also essentially the reason we can expect the resulting theories to be integrable; the scattering properties of the deformed theories are the same as those of the original.

When we deform the anti-de Sitter part of space-time there are certain caveats we should keep in mind, and we will address them below for the specific deformations. Nonetheless, in general deformations of $\mbox{AdS}_5$ have merit since they allow us to partly restore supersymmetry broken by deformations of the sphere, as well as providing a nice regularization of the ground state energy of the undeformed theory as we will discuss in section \ref{sec:GSobservations}. Moreover a completely general deformation breaks all symmetries of the original model, lifting the $\mbox{PSU}(2,2|4)$ degeneracy of the spectrum completely, giving the richest possible structure\footnote{Deformations also lift accidental degeneracies in the asymptotic spectrum \cite{arXiv:1103.2708,Sfondrini:2011rr}.}. In this section we will first discuss strings on orbifolds, followed by strings on TsT transformed backgrounds, and finally we will briefly summarize the link between these boundary conditions and elements of the bosonic symmetry group of the light-cone gauge fixed superstring at the level of the coset model.

\subsection{Orbifolds}

Starting with a (super-)string theory on (super-)space-time $\mathcal{M}$ and the action of a discrete group $\Gamma$ on $\mathcal{M}$, we can orbifold the original string theory by $\Gamma$ and obtain a string theory on $\mathcal{M}/\Gamma$ \cite{Dixon:1985jw}. The orbifolded background is described in terms of the fields $X \in \mathcal{M}$ of the original string theory as
\begin{equation}
X(\tau,\sigma) \simeq g X(\tau,\sigma) \, , \, \, g \in \Gamma \, . \label{eq:orbequiv}
\end{equation}
Were we considering point particles on this space, we would take the Hilbert space of the original model, and project it onto its $\Gamma$ invariant subspace. However in the case of strings we have more structure; because of the orbifold equivalence (\ref{eq:orbequiv}) the string does not have to close on itself. Rather it can also close modulo an orbifold group element
\begin{equation}
\label{eq:orbequivBC}
X(\tau,2 \pi ) \simeq h X(\tau,0) \, , \, \, h \in \Gamma \, .
\end{equation}
In the orbifolded theory, we need to consider all independent sectors of this type, which are known as twisted sectors. These twisted sectors are in one to one correspondence with the conjugacy classes of $\Gamma$, as follows from (\ref{eq:orbequiv},\ref{eq:orbequivBC}). As such, in each twisted sector we should project onto its $\Gamma - [g]$ invariant subspace, where $[g]$ is the corresponding conjugacy class. In other words, we should project on the subspace invariant under the stabilizer of $g$, $\Gamma_g$.

For the $\ads$ superstring, this means we want to consider strings on $\mbox{AdS}_5/\Gamma^a \times \mbox{S}^5 /\Gamma^s$ by following these considerations. Let us start by discussing the boundary conditions obtained by orbifolding the sphere.

\subsubsection*{Orbifolding the sphere - $S^5/\Gamma$}

The possible orbifolds of the sphere are obtained by the action of the discrete subgroups of $\mbox{SU}(4)$. Fortunately enough, these have been classified completely in \cite{Hanany:1999sp}. Of course, all elements of the abelian subgroups can be simultaneously diagonalized, and hence immediately put into correspondence with elements of the Cartan subgroup of $\mbox{SU}(4)$. This simply corresponds to one and the same field redefinition in every twisted sector; explicitly
\begin{equation}
X(\tau,\sigma) \rightarrow \tilde{X}(\tau,\sigma) = g X(\tau,\sigma)\, , \, \, g \in \mbox{SU}(4) \, ,
\end{equation}
such that each of the representatives $h_k$ of the conjugacy classes is diagonalized
\begin{equation}
\tilde{X}(\tau,2 \pi ) \simeq d_k \tilde{X}(\tau,0) \, ,
\end{equation}
where $d_k$ is in the Cartan subgroup of $\mbox{SU}(4)$. Simultaneous diagonalization is not possible within a non-abelian subgroup; still we can diagonalize the representative conjugacy class element in each twisted sector by an independent field redefinition, as already used in this context in \cite{Solovyov:2007pw}. Picking up a different basis in each twisted sector is not a problem since each sector is closed\footnote{This holds as long as the quantum orbifold symmetry is not broken. Non-supersymmetric orbifolds can break this symmetry and we come back to this point shortly.}. In short, in every twisted sector we can restrict ourselves to boundary conditions corresponding to elements of the Cartan subgroup of $\mbox{SU}(4)$; boundary conditions on the isometry angles of the sphere. Compared to abelian orbifolds the only complication for non-abelian orbifolds is in the bookkeeping of the twisted sectors and corresponding orbifold invariance as clearly explained in \cite{Solovyov:2007pw}\footnote{In \cite{Solovyov:2007pw} also the quiver structure of these orbifolds is carefully worked out.}.

The string theory constructed in this fashion is dual to an orbifolded version of $\mathcal{N}=4$ SYM, with field content $\mathcal{G}/\Gamma$ where $\mathcal{G}$ denotes the field content of $\mathcal{N}=4$ SYM. Correspondingly, $\Gamma$ is a discrete subgroup of the $R$-symmetry group of the field theory, meaning that these orbifolds will break a certain amount of supersymmetry. The amount of supersymmetry that is preserved after orbifolding can be found by considering the embedding of the orbifold in $\mbox{SU}(4)$, in other words finding the residual $R$-symmetry; we obtain an $\mathcal{N}=2$ theory if $\Gamma \subset \mbox{SU}(2)$, whereas $\Gamma \subset \mbox{SU}(3)$ gives us an $\mathcal{N}=1$ theory. Of course, taking $\Gamma \subset \mbox{SU}(4)$ gives a non-supersymmetric theory.

On the string theory side such a non-supersymmetric theory suffers from instabilities \cite{Tseytlin:1999ii,Dymarsky:2005uh} related to tachyon condensation. The corresponding vacuum expectation values of the dual twisted operators break the quantum orbifold symmetry \cite{Adams:2001jb,Dymarsky:2005nc}\footnote{This should happen at intermediate values of the coupling; at large coupling there are no tachyons in the spectrum.}. Apart from the instabilities themselves, this also presents us with a practical problem; without orbifold symmetry the Hilbert space of the theory no longer necessarily decomposes over the various twisted sectors. For non-abelian orbifolds this would greatly complicate the problem by preventing us from choosing a different field basis in each twisted sector. We might wish to disregard such theories and focus on orbifolds with $\Gamma \subset \mbox{SU}(3)$ at most. However, there are no practical obstacles in studying non-supersymmetric abelian orbifolds so we will consider them below. In line with this, non-supersymmetric abelian orbifolds of $\mathcal{N} = 4$ SYM have been considered before \cite{Beisert:2005he}, while non-abelian ones have not \cite{Solovyov:2007pw}. As we already mentioned, in certain cases part of the supersymmetry can be restored by simultaneously orbifolding $\mbox{AdS}_5$.

\subsubsection*{Orbifolding anti-de Sitter space - $\mbox{AdS}_5/\Gamma$}

Exactly as we did for the sphere we can consider orbifolds of $\mbox{AdS}_5$. However, since $\mbox{AdS}_5$ contains the time-like direction of our background space-time we should proceed cautiously to avoid possible time-like orbifolds. With this restriction in mind, we are left to consider orbifolds by $\Gamma \subset \mbox{SU}(2)\times \mbox{SU}(2) \subset \mbox{SU}(2,2)$. For these orbifolds we can proceed as we did for the sphere, again reducing the boundary conditions to those corresponding to diagonal elements of $\mbox{SU}(2)\times \mbox{SU}(2)$ in each twisted sector. Rather than breaking supersymmetry, now we break part of the conformal symmetry. From the point of view of integrability of the string theory this is no real objection as we will see below\footnote{Formally as far as integrability is concerned we could even consider time-like orbifolds of $\mbox{AdS}_5$, but the corresponding spectral problem is not very exciting.}. While the details of the corresponding dual field theories have not been worked out in full detail, at the level of the Bethe ansatz describing the spectrum of the dilatation operator such orbifolds can also be readily accounted for \cite{Beisert:2005he}. Naturally it is possible to simultaneously orbifold $\mbox{AdS}_5$ and $\mbox{S}^5$ in any of the ways we discussed.

\subsection{TsT transformed backgrounds}

The second type of deformation we can apply to our background is a sequence of so-called TsT transformations which can be applied to backgrounds with more than one commuting $\mbox{U}(1)$ isometries. A TsT transformation takes a pair of angles corresponding to two isometry directions, say $(\phi_1, \phi_2)$, and acts on them via a $T$-duality along the $\phi_1$ direction, followed by a shift in the $\phi_2$ direction\footnote{This variable is not affected by the $T$-duality.}, $\phi_2 \rightarrow \phi_2 + \gamma \tilde{\phi}_1$, and finally $T$-dualizing back in the $\tilde{\phi}_1$ direction, where $\tilde{\phi}_1$ is the $T$-dual variable to $\phi_1$. By sequentially applying TsT transformations on a background with $d$ commuting $\mbox{U}(1)$ isometries we can obtain a $d(d-1)/2$ parametric deformation. Note that these transformations require all other fields to be uncharged under the relevant $\mbox{U}(1)$s, specifically in this context the fermions. Fortunately they can be discharged by a field redefinition \cite{hep-th/0512253}.

Very important for our present considerations, the resulting equations of motion are in one-to-one correspondence with the untransformed equations of motion with twisted boundary conditions imposed on the $\mbox{U}(1)$ isometry fields \cite{hep-th/0503201}. In fact, since by definition these $\mbox{U}(1)$ isometries are nothing but elements of the Cartan subgroup of the isometry group of $\ads$, it should not be surprising that the resulting boundary conditions can be put into correspondence with elements of the Cartan subgroup\footnote{The fact that this works also for fermions is the result of the non-trivial field redefinition of \cite{hep-th/0512253}.}, exactly as we could for orbifolds. Let us now briefly discuss the various possible TsT transformations.

Deformations of the sphere are very sensible from the point of view of AdS/CFT. Restricting ourselves to the sphere a generic sequence of TsT transformations gives a three parameter deformed background \cite{hep-th/0503201}, known as a $\gamma$-deformed background. This background contains the Lunin-Maldacena background \cite{hep-th/0502086} for real $\beta$ as a specific case. The corresponding dual field theories in general have no supersymmetry, with the exception of the Lunin-Maldacena case which preserves $\mathcal{N} =1$ supersymmetry. To describe string propagation on this background through the original theory, we have to implement the following boundary conditions on the three angles on the sphere
\begin{equation}
\phi_i(2\pi) - \phi_i(0) = 2\pi(n_i -\epsilon_{ijk}\gamma_j J_k)\, , \, \, \, n_i \in \mathbb{Z}\, .
\end{equation}
Here the $J_k$ are the angular momenta of $S^5$ and $\gamma_j$ the three parameters of the deformation.

\subsubsection*{TsT transformations involving $\mbox{AdS}_5$}

Less studied but clearly possible from the point of view of string theory, we can do TsT transformations on the three commuting $\mbox{U}(1)$ isometries of anti-de Sitter space \cite{hep-th/0605018,Swanson:2007dh}. However, as always we should be proceed cautiously when time-like directions are involved. Also, while they give sensible string backgrounds these deformations have a less clear interpretation from the point of view of AdS/CFT.

Let us briefly address the issues regarding TsT transformations involving $\mbox{AdS}_5$, starting with issues regarding TsT transformations involving time-like directions. First of all, strictly speaking our string theory is defined on the universal cover of $\mbox{AdS}_5$ which does not have a compact time direction, such that $T$-dualizing in this direction is not an option. Secondly, if we do a TsT transformation involving the time direction we will generate space-time directions with mixed signature. To circumvent the first problem we could initially consider TsT transformations on the space-time, and only subsequently pass to its universal cover. Also, the second problem can pragmatically be avoided by restricting study of the corresponding string theory to regions of the geometry where the signature is the same as the parent theory \cite{hep-th/0605018}. In fact, we could proceed in the spirit of \cite{hep-th/0605018} and formally apply a general sequence of TsT transformations, whereby we would not necessarily view the resulting geometry as a deformation of any particular parent geometry, rather studying it at face value.

Furthermore, any TsT transformation involving an angle from $\mbox{AdS}_5$ necessarily breaks part of the conformal symmetry. The corresponding free string theory will be sensible, even integrable, but again the effect of breaking conformal symmetry on the dual field theory is not obvious. By analogy to how $\beta$-deformed SYM can be obtained by introducing a $\star$-product related to the deformation in the $R$-symmetry group the duals are expected to be certain non-commutative dual field theories \cite{Beisert:2005if,hep-th/0605018}. Despite these difficulties, TsT transformations of this type have been investigated in the setting of AdS/CFT. In fact, as far as integrability is concerned we can readily implement any of these deformations, so while keeping the above concerns in mind we will discuss generic TsT transformations of $\ads$. As we can expect the interesting features of these deformations turn to rather peculiar ones as soon as time-like directions are involved.

The boundary conditions corresponding to the most generic sequence of TsT transformations are given by
\begin{equation}\label{eqn;genTsT}
\phi_i (2\pi) -\phi_i(0) =  2\pi(n_i + \gamma_{ik} J_k) \, , \, \, \, n_i \in \mathbb{Z}\,
\end{equation}
where $i$ is a generalized index labeling the six $\mbox{U}(1)$ isometry fields of $\ads$, $J_i$ the corresponding angular momentum and $\gamma_{ij}$ an antisymmetric six by six matrix containing the $15$ deformations parameters. Clearly we can also combine orbifolds and TsT transformations \cite{Beisert:2005he}.

\subsection{Deformations in the light-cone gauge}

We have seen that strings on both orbifolds of, and TsT transformed $\ads$ can be directly described through the original string theory with modified boundary conditions. Let us parametrize a completely generic deformation of the above type by the boundary conditions
\begin{equation}
\label{eq:genBCs}
\psi_i (2\pi) = \psi_i (0) + \alpha_i \, , \, \, \, \phi_j (2\pi) = \phi_j (0) + \beta_i \,,
\end{equation}
where $\psi_1 \equiv t$ and $\phi_1 \equiv \phi$ with associated Noether charges $E$ and $J$ respectively, and $\psi_{2,3}$ and $\phi_{2,3}$ are the remaining $\mbox{U}(1)$ isometry fields in $\mbox{AdS}_5$ and $\mbox{S}^5$ respectively. The boundary condition on $t$ in the form of $\alpha_1$ is a formal bookkeeping device and not meant to be taken in a direct sense. We introduce it for uniformity since it can appear for general TsT transformed backgrounds \textit{cf.} \eqref{eqn;genTsT} with all their caveats, for orbifolds it is absent. We would like to explicitly mention here that the parameters $\alpha_i$ and $\beta_i$ can be taken between zero and $2 \pi$ since only their values modulo $2 \pi$ have physical relevance.

Up to now the boundary conditions on the isometry fields were all treated on an equal footing, but this changes when fixing a light-cone gauge. Fixing a light-cone gauge breaks the manifest $\mbox{SU(2,2)} \times \mbox{SU}(4)$ bosonic symmetry of the superstring to $\mbox{SU}(2)^4$ by fixing the combination $x^+ \equiv t +\phi$ to be equal to the proper time parametrizing the propagation of the superstring. Naturally, the information on the non-trivial boundary conditions on $t$ and $\phi$ is far from lost; we get the so-called level matching condition from the boundary conditions on $x^- \equiv t -\phi$
\begin{equation}
\label{eq:pwslevelmatching}
p_{ws} = \int_0^{2\pi} d \sigma x^{-\prime} = \alpha_1 - \beta_1 + 2 \pi m \, ,
\end{equation}
where $m$ is the winding number of the string\footnote{This of course nicely matches with the condition $e^{i P} =1$ for the total momentum of the undeformed SYM spin chain.}. We would like to emphasize that $p_{ws}$ is the world-sheet momentum of the light-cone gauge fixed superstring with quasi-periodic boundary conditions, not of the superstring on a deformed background. Deformations in the $x^+$ direction are simply world sheet diffeomorphisms. Let us also recall that the fermionic fields are not necessarily periodic, but rather have boundary conditions related to the level matching condition, \textit{i.e.}
\begin{equation}
\eta(2\pi) = e^{-\frac{i}{2} p_{ws}} \eta(0) \,, \, \, \, \theta(2\pi) = e^{-\frac{i}{2} p_{ws}} \theta(0) \, .
\end{equation}
Here $\eta$ and $\theta$ denote the fermionic fields of the light-cone gauge fixed coset model \cite{Arutyunov:2009ga}. In particular this means that the fermions of the undeformed model are anti-periodic in odd winding sectors. The remaining boundary conditions corresponding to $\alpha_{2,3}$ and $\beta_{2,3}$ can now be put in correspondence with the four remaining Cartan generators of $\mbox{SU}(2)^4$. The easiest way to make this link is to consider the generators of shifts of the angles $\psi_{2,3}$ and $\phi_{2,3}$ in the fundamental of $\mbox{SU}(2,2)$ and $\mbox{SU}(4)$ \cite{Arutyunov:2009ga}
\begin{equation}
C_2 = \mbox{diag}(-1,1,-1,1) \, , \, \, \, C_3 = \mbox{diag}(-1,1,1,-1)
\end{equation}
and the diagonal embedding of the Cartan generators of the four unbroken $\mbox{SU}(2)$s in $\mbox{SU}(2,2)$ and $\mbox{SU}(4)$, which are each just the usual
\begin{equation}
C = \mbox{diag}(1,-1) \, .
\end{equation}
If we then parametrize the four diagonal elements of $\mbox{SU}(2)$ by $\alpha$ and $\dot{\alpha}$ for anti-de Sitter and $\beta$ and $\dot{\beta}$ for the sphere we get
\begin{align}
\alpha & = -\frac{\alpha_2+\alpha_3}{2} \, , \, \, \, \dot{\alpha} = \frac{\alpha_3-\alpha_2}{2} \, , \label{gentwist1}\\
\beta & = -\frac{\beta_2+\beta_3}{2} \, , \, \, \, \dot{\beta} = \frac{\beta_3-\beta_2}{2} \,
\label{gentwist2}.
\end{align}
Let us introduce $\gamma \equiv \alpha_1-\beta_1$ to parametrize the level matching condition \eqref{eq:pwslevelmatching}. Now we are ready to implement these boundary conditions in the transfer matrix of our integrable model.

\section{Twisted Transfer Matrices}
\label{sec:TTM}

A important object in integrable models is the transfer matrix. As it turns out the transfer matrix of a theory with periodic boundary conditions has a natural generalization to the case of quasi-periodic boundary conditions \cite{E-11-86}, which is exactly what our deformations correspond to. In this section we discuss the different types of integrable twists of the $\ads$ superstring on the level of the transfer matrix. We will first discuss transfer matrices and twists in general before specializing to the $\ads$ superstring.

\subsection{General theory}

\subsubsection*{The S-matrix and the transfer matrix}

Let us consider $K$ particles in an integrable model with momenta $p_1,\ldots,p_{K}$ transforming in some representation of a symmetry algebra and scattering with a corresponding S-matrix $\S(p_i,p_j)$. To these particles we add an auxiliary one with momentum $q$. Any state of this system then lives in the following tensor product space
\begin{eqnarray}
\mathcal{V}:=V_0(q)\otimes V_1(p_1)\otimes \ldots \otimes V_K(p_{K}),
\end{eqnarray}
where $V_i$ is a carrier space of the representation of the $i$-th particle. It is instructive to split the states into an auxiliary piece and a physical piece
\begin{eqnarray}
|A\rangle_0\otimes|B\rangle_{K} \in \mathcal{V},\nonumber
\end{eqnarray}
where $|A\rangle_0 \in V(q)$ and \footnote{All the tensor products are defined with increasing order of the index as $1,2,\ldots, K$.} $|B\rangle_K \in  V_P:=\bigotimes_{i} V(p_{i})$. The monodromy matrix acting in the space $\mathcal{V}$ is then defined as follows
\begin{eqnarray}
\mathcal{T}_0(q|\vec{p}) :=
\prod_{i=1}^{K^{\rm{I}}}\mathbb{S}_{0k}(q,p_i),
\end{eqnarray}
where $\mathbb{S}_{0k}(q,p_k)$ is the S-matrix describing the scattering between the auxiliary particle and a `physical' particle with momentum $p_k$. This operator manifestly depends on the representation of the auxiliary particle.

The monodromy matrix can be seen as a matrix in the auxiliary space $V_0(q)$,  the corresponding matrix elements  being themselves operators on $V_P$. Indeed, introducing a basis $|e_{I}\rangle$ for $V_0(q)$ and a basis $|f_{A}\rangle$ for $V_P$, the action of the monodromy matrix $\mathcal{T}\equiv \mathcal{T}_0(q|\vec{p})$ on the total space $\mathcal{V}$ can be written as
\begin{eqnarray}\label{eqn;ActionMonodromy}
\mathcal{T}(|e_{I}\rangle\otimes |f_A\rangle) = \sum_{J,B}
T_{IA}^{JB}|e_{J}\rangle\otimes |f_B\rangle.
\end{eqnarray}
The matrix entries of the monodromy matrix can then accordingly be denoted as
\begin{eqnarray}
\mathcal{T}|e_{I}\rangle = \sum_{J} \mathcal{T}^{J}_{I}
|e_{J}\rangle\,  ,
\end{eqnarray}
while the action of the matrix elements $\mathcal{T}^{J}_{I}$ as
operators on $V_P$ can easily be read off as
\begin{eqnarray}
\mathcal{T}^{J}_{I}|f_A\rangle = \sum_{B} T_{IA}^{JB}|f_B\rangle.
\end{eqnarray}
The operators $\mathcal{T}^{J}_{I}$ have non-trivial commutation relations among themselves. Consider two different auxiliary spaces $V_a(q),V_b(\tilde{q})$. The Yang-Baxter equation for $\S$ implies that
\begin{align}\label{eqn;YBE-operators}
\mathbb{S}_{ab} (q,\tilde{q})\,\mathcal{T}_a(q|\vec{p})\mathcal{T}_b(\tilde{q}|\vec{p}) = \mathcal{T}_b(\tilde{q}|\vec{p})\mathcal{T}_a(q|\vec{p})\,\mathbb{S}_{ab} (q,\tilde{q}),
\end{align}
where $\mathbb{S}(q,\tilde{q})$ is the S-matrix describing the scattering between the two auxiliary particles. By explicitly working out these relations, we can find the commutation relations between the different matrix elements of the monodromy matrix. These so-called fundamental commutation relations (\ref{eqn;YBE-operators}) constitute the cornerstone of the Algebraic Bethe Ansatz \cite{Faddeev:1996iy}.

The transfer matrix is subsequently defined as
\begin{eqnarray}
T_0(q|\vec{p}):={\rm{str}}_0\mathcal{T}_{0}(q|\vec{p}),
\end{eqnarray}
and it can be viewed as an operator acting on the physical space $V_P$. By virtue of the Yang-Baxter equation it defines an infinite set of commuting quantities
\begin{align}
[T(q),T(\tilde{q})] =0,
\end{align}
making the integrability of the model explicit.

\subsubsection*{Integrable twists}

Let us again consider a transfer matrix $T = \mathrm{str}_a S_{a1}\ldots S_{aK}$ build up out of an S-matrix that commutes with elements $g\in G$ from a group $G$
\begin{align}
[\S,g\otimes g] = 0.
\end{align}
We can then define a twisted transfer matrix $T^{g}$ as follows
\begin{align}
 T^{g} = \mathrm{str}_a \, g_a S_{a1} \ldots S_{aK}.
\end{align}
These twists are integrable as they preserve the crucial property of integrability that $[ T^{g}(p), T^{g}(q)]=0$, which follows straightforwardly from the Yang-Baxter equation. As already mentioned such a twisted transfer matrix actually describes the integrable model with quasi-periodic boundary conditions determined by the twist $g$ \cite{E-11-86}.

\subsection{Transfer matrix for $\ads$}

The S-matrix describing the scattering of the $\ads$ superstring consists of two copies of the centrally extended $\alg{su}(2|2)$ invariant S-matrix. The transfer matrix is the product of two transfer matrices each based on a $\mathfrak{su}(2|2)$ invariant S-matrix.

\subsubsection*{Allowed twists}

The centrally extended $\mathfrak{su}(2|2)$ invariant S-matrix commutes with elements from $SU(2)^2$, which is simply half of the residual symmetry of the superstring in the light-cone gauge. More precisely
\begin{align}\label{eqn;Scomm}
&[\S,G\otimes G] =0, && G\in SU(2)\times SU(2).
\end{align}
For any element $G\in SU(2)\times SU(2)$ we then define the following twisted transfer matrix
\begin{align}\label{eq:twistedtfgeneral}
T^G(q) = \mathrm{str}_a\,G_a\,\S_{a1}(q,p_1)\,\ldots\,\S_{aK}(q,p_K).
\end{align}
We will see below that for TsT transformations our twist element $G$ depends on the excitation numbers of the state under consideration so that in general $G$ acts also in the quantum space. Within each eigenspace however, the action of $G$ is purely in the auxiliary space. Rather than considering a generic element from $SU(2)^2$ we can reduce the types of twist we need to consider. It is not difficult to show that any element $G_1\in SU(2)$ can be written as
\begin{align}
&G_1 = U H U^{-1}, && U \in SU(2),
\,H = \begin{pmatrix} e^{i\phi} &0 \\ 0 & e^{-i\phi}\end{pmatrix} \in SU(2).
\end{align}
Since $U\in SU(2)$ and we have (\ref{eqn;Scomm}) it is easy to see that
\begin{align}
&T^G(q) = \vec{U}^{-1} \, T^H(q)\, \vec{U}, && \vec{U} = U_1\ldots U_K.
\end{align}
In other words, the transfer matrices are related via a simple basis transformation on the physical space. Consequently, the eigenvalues of $T^G$ and $T^H$ coincide and we will from now on just restrict to diagonal group elements, {\it i.e.} we will consider a group element of the form
\begin{align}\label{eqn;twist}
G = \mathrm{diag}(e^{i\alpha},e^{-i\alpha},e^{i\beta},e^{-i\beta})\, \in SU(2)\times SU(2),
\end{align}
where at the geometric level $\alpha$ affects the boundary conditions on $\mathrm{AdS}_5$ and $\beta$ those on $\mathrm{S}^5$, \textit{cf.} (\ref{gentwist1},\ref{gentwist2}).

For the transfer matrices that are used in the TBA approach we need the explicit form of $G$ in bound state representations. These representations are realized by monomials in superspace with bosonic variables $w_{1,2}$ and their fermionic counterparts $\theta_{3,4}$ \cite{arXiv:0803.4323}. The $\ell$ particle bound state representation is then spanned by the monomials of degree $\ell$ and the symmetry generators are given by differential operators. In this formalism our twist $G$ has the following explicit action on the states
\begin{align}\label{eqn;twistaction}
G(\theta_3^m\theta_4^n w_1^k w_2^l) = e^{i\alpha(k-l)+i\beta(m-n)}\theta_3^m\theta_4^n w_1^k w_2^l.
\end{align}
Since $m,n$ only take the values $0,1$ the $\beta$ dependent part is rather trivial, whereas the $\alpha$ dependent piece depends more intricately on the state. For concreteness, let us introduce the basis for an $\ell_0$-particle bound state representation as in \cite{Arutyunov:2009iq}
\begin{align}\label{basisV0}
&e_{\alpha;k} := \theta_{\alpha} w_1^{\ell_0-k-1}w_{2}^{k},
&&e_{k} := w_1^{\ell_0-k}w_{2}^{k},
&&e_{34;k} := \theta_{3}\theta_{4} w_1^{\ell_0-k-1}w_{2}^{k-1}.
\end{align}
On the level of the transfer matrix, these coefficients present themselves as relative weights in the sum
\begin{align}\label{eqn;Transfer}
T^{G}(q)=\sum_{k=0}^{\ell_0} e^{i\alpha(\ell_0-2k)}\mathcal{T}^{k}_{k} + \sum_{k=1}^{\ell_0-1}  e^{i\alpha(\ell_0-2k)} \mathcal{T}^{34;k}_{34;k} -
\sum_{k=0}^{\ell_0-1} \, e^{i\alpha(\ell_0-2k-1)}\left[e^{i\beta}\mathcal{T}^{3;k}_{3;k} + e^{-i\beta}\mathcal{T}^{4;k}_{4;k}\right],
\end{align}
where the auxiliary particle is a $\ell_0$ particle bound state. The twist can be seen as a rescaling of the elements of the monodromy matrix.

\subsubsection*{The $\mathfrak{su}(2)$ vacuum}

Consider the fermionic vacuum with all physical particles in the fundamental representation
\begin{eqnarray}
|0\rangle_P^{\prime}= \theta_{3}\otimes\ldots\otimes\theta_{3}.
\end{eqnarray}
It is easy to check that this vacuum is an eigenstate of the twisted transfer matrix. From the explicit form of the bound state S-matrix \cite{arXiv:0902.0183}, the action of the diagonal elements of fermionic type of the monodromy matrix is given by
\begin{align}\label{frmi}
&\mathcal{T}_{3;k}^{3;k}|0\rangle_P^{\prime} =
\prod_{i=1}^{K^{\rm{I}}}\frac{x_0^--x_i^+}{x^+_0-x^-_i}\sqrt{\frac{x^+_0x^-_i}{x^-_0x^+_i}}|0\rangle_P^{\prime},&&
\mathcal{T}_{4;k}^{4;k}|0\rangle_P^{\prime} =
\prod_{i=1}^{K^{\rm{I}}}\frac{x_0^--x_i^-}{x^+_0-x^-_i}\frac{x_i^--
\frac{1}{x_0^+}}{x_i^+-\frac{1}{x_0^+}}\sqrt{\frac{x^+_0x^+_i}{x^-_0x^-_i}}|0\rangle_P^{\prime}.
\end{align}
Notice that these elements are \emph{independent} of $k$.

The next step is to consider the bosonic elements $\mathcal{T}^k_k,\mathcal{T}^{34,k}_{34,k}$. Also these elements act in a straightforward way
\begin{eqnarray}
\mathcal{T}^k_k|0\rangle_P^{\prime}&=&\prod_{i=1}^{K^{\rm{I}}}\frac{x_0^--x^-_i}{x_0^+-x^-_i}\sqrt{\frac{x^+_0}{x^-_0}}|0\rangle_P^{\prime}
\end{eqnarray}
and
\begin{eqnarray}
\mathcal{T}^{\alpha4,k}_{\alpha4,k}|0\rangle_P^{\prime}&=&\prod_{i=1}^{K^{\rm{I}}}\frac{x_0^--x^+_i}{x_0^+-x^-_i}\frac{x_i^--\frac{1}{x_0^+}}{x_i^+-\frac{1}{x_0^+}}\sqrt{\frac{x^+_0}{x^-_0}}|0\rangle_P^{\prime}.
\end{eqnarray}
Similarly to the fermionic contributions (\ref{frmi}), we once again find that these terms are \emph{independent} of $k$.

The twist (\ref{eqn;twist}) acts straightforwardly on the elements via (\ref{eqn;Transfer}). Summing everything is straightforward due to the trivial $k$ dependence. This finally gives that $|0\rangle_P^{\prime}$ is an eigenvalue of the transfer matrix with eigenvalue
\begin{align}
\label{anti}
\Lambda(q|\vec{p})=&\frac{\sin (\ell_0+1)\alpha}{\sin\alpha}\prod_{i=1}^{K^{\rm{I}}}\frac{x_0^--x^-_i}{x_0^+-x^-_i}
\sqrt{\frac{x^+_0}{x^-_0}}
+\frac{\sin (\ell_0-1)\alpha}{\sin\alpha}\prod_{i=1}^{K^{\rm{I}}}\frac{x_0^--x^+_i}{x_0^+-x^-_i}
\frac{x_i^--\frac{1}{x_0^+}}{x_i^+-\frac{1}{x_0^+}}\sqrt{\frac{x^+_0}{x^-_0}}-
\nonumber \\
&-\frac{\sin \ell_0\,\alpha}{\sin\alpha}
\left[e^{i\beta}\prod_{i=1}^{K^{\rm{I}}}\frac{x_0^--x_i^+}{x^+_0-x^-_i}
\sqrt{\frac{x^+_0x^-_i}{x^-_0x^+_i}}
+e^{-i\beta}
\prod_{i=1}^{K^{\rm{I}}}\frac{x_0^--x_i^-}{x^+_0-x^-_i}\frac{x_i^--\frac{1}{x_0^+}}
{x_i^+-\frac{1}{x_0^+}}\sqrt{\frac{x^+_0x^+_i}{x^-_0x^-_i}}\right].
\end{align}
This precisely agrees with the result of \cite{Beisert:2006qh} for antisymmetric representations in the limit where the twists vanish.

\subsubsection*{The $\mathfrak{sl}(2)$ transfer matrix}

Just as for the $\mathfrak{su}(2)$ vacuum discussed above, the twist modifies the eigenvalue of the vacuum $|0\rangle_P = \otimes_i w_1^{\ell_i}$ in a straightforward way compared to \cite{Arutyunov:2009iq}. Introducing the following functions
\begin{eqnarray}\label{eqn;lambda-pm}
\lambda_\pm(q,p_i,k)&=&\frac{\mathscr{X}^{k,0}_k}{2\mathcal{D}}\left[1-\frac{(x^-_ix^+_0-1)
   (x^+_0-x^+_i)}{(x^-_i-x^+_0)
   (x^+_0x^+_i-1)}+\frac{2ik}{g}\frac{x^+_0
   (x^-_i+x^+_i)}{(x^-_i-x^+_0)
   (x^+_0x^+_i-1)}\right.\\
&&\left.\pm\frac{i x^+_0
   (x^-_i-x^+_i)}{(x^-_i-x^+_0)
 (x^+_0x^+_i-1)}\sqrt{\left(\frac{2k}{g}\right)^2+2i
\left[x^+_0+\frac{1}{x^+_0}\right]
\frac{2k}{g}-\left[x^+_0-\frac{1}{x^+_0}\right]^2}\right]\nonumber,
\end{eqnarray}
we find that the vacuum eigenvalue is given by
\begin{eqnarray}\label{transfer-rankone}
\Lambda(q|\vec{p})&=&e^{i\ell_0 \alpha} + e^{-i\ell_0 \alpha}\prod_{i=1}^{K^{\rm{I}}}
\left[\frac{(x^-_0-x^-_i)(1-x^-_0 x^+_i)}{(x^-_0-x^+_i)(1-x^+_0
x^+_i)}\sqrt{\frac{x^+_0x^+_i}{x^-_0x^-_i}}\mathscr{X}^{\ell_0,0}_{\ell_0}\right]\nonumber\\
&&-(e^{i\beta}+e^{-i\beta})\sum_{k=0}^{\ell_0-1}e^{i\alpha(\ell_0-2k-1)}\prod_{i=1}^{K^{\rm{I}}}\left(\frac{x^+_0-x^+_i}{x^-_0-x^+_i}
\sqrt{\frac{x^-_0}{x^+_0}}\left[1-\frac{k}{u_0-u_i+\frac{\ell_0-\ell_i}{2}
}\right]\mathscr{X}^{k,0}_k\right) \nonumber\\
&& +
\sum_{k=1}^{\ell_0-1}e^{i\alpha(\ell_0-2k)}\left[\prod_{i=1}^{K^{\rm{I}}}\lambda_+(q,p_i,k)
+\prod_{i=1}^{K^{\rm{I}}}\lambda_-(q,p_i,k)\right].
\end{eqnarray}
For the fundamental case ($\ell_0=\ell_i=1 \, \, \forall i$), this reduces to
\begin{eqnarray}\label{T fundamental-all}\nonumber
T_0(q|\vec{p})|0\rangle_P
&=&\left\{e^{i\alpha} + e^{-i\alpha}\prod_{i=1}^{K^{\rm{I}}}
\frac{x^+_0-x^+_i}{x^+_0-x^-_i}\frac{1-\frac{1}{x^-_0x^+_i}}{1-\frac{1}{x^-_0x^-_i}}
-(e^{i\beta}+e^{-i\beta})
\prod_{i=1}^{K^{\rm{I}}}\frac{x^+_0-x^+_i}{x^+_0-x^-_i}\sqrt{\frac{x^-_i}{x^+_i}}\right\}|0\rangle_P.
\end{eqnarray}
Following \cite{Arutyunov:2009iq}, we continue with the construction of excited states by considering creation operators $B_\alpha$ depending on auxiliary rapidities. Obviously, by only twisting the transfer matrix the fundamental commutation relations are not altered. Thus, the commutation relation of $B$ with the bosonic elements from $\mathcal{T}$ are of the form
\begin{align}
(T^k_k + T_{34;k}^{34;k}) B_\alpha \sim B_\alpha (T^k_k + T_{34;k}^{34;k}) ~+~ \mathrm{unwanted},
\end{align}
see \cite{Arutyunov:2009iq} for the exact relation. From this we see that all the relevant terms in this commutation relation are multiplied by the same factor in (\ref{eqn;Transfer}). In other words, the bosonic part of the twisted transfer matrix behaves trivially under the twist. The fermionic piece however, involves a non-trivial mixing
\begin{align}
T^{\gamma;k}_{\alpha_1,k} B_{\alpha_2} \sim B_{\beta_2} T^{\gamma;k}_{\beta_1} r^{\beta_1\beta_2}_{\alpha_1\alpha_2} ~ +~ \mathrm{unwanted},
\end{align}
where $r$ is the XXX R-matrix. This commutation relation mixes the two species of fermions, which are affected differently by the $\beta$-part of the twist. This translates into considering a {\it twisted} XXX spin chain as a nested system rather than the untwisted one as in \cite{Martins:2007hb,arXiv:0902.0183}. For detail on the twisted XXX spin chain we refer to appendix \ref{app:XXX}.

This finally results in the following transfer matrix for an anti-symmetric bound state representation with the bound state number $a$
\begin{align}\label{eqn;FullEignvalue}
T_{a,1}(v\,|\,\vec{u}) =&
\left.\prod_{i=1}^{K^{\rm{II}}} {\textstyle{\frac{y_i-x^-}{y_i-x^+} \sqrt{\frac{x^+}{x^-}} }}\right[
{\textstyle{e^{ia\alpha} + e^{-ia\alpha}}} \prod_{i=1}^{K^{\rm{II}}} {\textstyle{\frac{v - \nu_i + \frac{ia }{g} }{v-\nu_i-\frac{ia }{g}}}}
\prod_{i=1}^{K^{\rm{I}}}{\textstyle{\frac{(x^--x^-_i)(1-x^- x^+_i)x^+}{(x^+-x^-_i)(1-x^+ x^+_i)x^-}+ }}\nonumber\\
&{\textstyle{+}}\sum_{k=1}^{a-1}e^{i(a-2k)\alpha} \prod_{i=1}^{K^{\rm{II}}}
{\textstyle{\frac{v-\nu_i + \frac{ia }{g} }{v - \nu_i + \frac{i(a-2k)}{g}}}}
\left\{\prod_{i=1}^{K^{\rm{I}}}{\textstyle{\lambda_+(v,u_i,k)+}}\right.\left.\prod_{i=1}^{K^{\rm{I}}}{\textstyle{\lambda_-(v,u_i,k)}}\right\}\\
& - \sum_{k=0}^{a-1}e^{i(a-2k-1)\alpha}\prod_{i=1}^{K^{\rm{II}}} {\textstyle{\frac{v - \nu_i + \frac{ia }{g} }{v-\nu_i + \frac{i(a-2k)}{g}}}}
\prod_{i=1}^{K^{\rm{I}}}{\textstyle{\frac{x^+-x^+_i}{x^+-x^-_i}\sqrt{\frac{x^-_i}{x^+_i}} \left[1-\frac{\frac{2ik}{g}}{v-u_i+\frac{i(a-1)}{g}}\right]}}\times\nonumber\\
&\times\left. \left\{e^{i\beta}\prod_{i=1}^{K^{\rm{III}}}{\textstyle{\frac{w_i-v - \frac{i(a-2k+1)}{g}}{w_i- v - \frac{i(a-2k-1)}{g}} + }}
e^{-i\beta}\prod_{i=1}^{K^{\rm{II}}}{\textstyle{\frac{v-\nu_i + \frac{i(a-2k)}{g}}{v-\nu_i+\frac{i(a-2k-2)}{g}}}}
\prod_{i=1}^{K^{\rm{III}}}{\textstyle{\frac{w_i - v - \frac{i(a-2k-3)}{g}}{w_i-v - \frac{i(a-2k-1)}{g}}}}\right\}\right],\nonumber
\end{align}
provided the auxiliary roots $y, w$ satisfy the Bethe equations (\ref{eqn;FullBAE2}),(\ref{eqn;FullBAE3}). The rapidities $\nu$ are related to $y$ via $\nu = y+1/y$. Sending the twist parameters $\alpha,\beta$ to zero trivially reproduces the regular transfer matrix \cite{Martins:2007hb,arXiv:0902.0183}. The untwisted transfer matrix can also be computed from a generating functional \cite{Beisert:2006qh} as explicitly confirmed in \cite{arXiv:0902.0183}. This approach has also been applied to twisted transfer matrices and we briefly discuss this in appendix \ref{app:genfunctional}.

\subsection{Bethe equations}\label{sec:BAE}

The complete S-matrix consists of two copies of the $\mathfrak{su}(2|2)$ invariant S-matrix. We will refer to these copies as dotted and undotted respectively. Both the dotted and undotted copies are twisted with their own group element $G$ and we will correspondingly denote the parameters by $\alpha,\dot{\alpha}$ and $\beta,\dot{\beta}$. The main Bethe equation can be readily derived from the transfer matrix and is given by
\begin{align}\label{eqn;FullBAE}
1&= e^{i(\alpha+\dot{\alpha})\ell_k}\left(\frac{x_k^+}{x_k^-}\right)^J\prod_{l=1,l\neq k}^{K^{\rm{I}}}S_{0}(p_{k},p_{l})^2~
\prod_{l=1}^{K^{\rm{II}}}\frac{{x_{k}^{-}-y_{l}}}{x_{k}^{+}-y_{l}}\sqrt{\frac{x^+_k}{x^-_k}}~ \prod_{l=1}^{\dot{K}^{\rm{II}}}\frac{{x_{k}^{-}-\dot{y}_{l}}}{x_{k}^{+}-\dot{y}_{l}}\sqrt{\frac{x^+_k}{x^-_k}}.
\end{align}
It is supplemented by the auxiliary Bethe equations
\begin{align}
1&=e^{i(\beta - \alpha)}\prod_{l=1}^{K^{\mathrm{I}}}\frac{y_{k}-x^{+}_{l}}{y_{k}-x^{-}_{l}}\sqrt{\frac{x^-_k}{x^+_k}}
\prod_{l=1}^{K^{\mathrm{III}}}\frac{y_{k}+\frac{1}{y_{k}}-w_{l}+\frac{i}{g}}{y_{k}+\frac{1}{y_{k}}- w_{l}-\frac{i}{g}}\label{eqn;FullBAE2}\\
1&=e^{-2i\beta}\prod_{l=1}^{K^{\mathrm{II}}}\frac{w_{k}-y_{l}-\frac{1}{y_{l}}+\frac{i}{g}}{w_{k}-y_{l}-\frac{1}{y_{l}}-\frac{i}{g}}
\prod_{l\neq
k}^{K^{\mathrm{III}}}\frac{w_{k}-w_{l}-\frac{2i}{g}}{w_{k}-w_{l}+\frac{2i}{g}}\label{eqn;FullBAE3},
\end{align}
together with their dotted counterparts. The parameters are related to the momentum and coupling constant via the usual constraints
\begin{align}
&x_{k}^{+}+\frac{1}{x_{k}^+}-x_{k}^{-}-\frac{1}{x_{k}^-} = \frac{2i
\ell_{k}}{g},&&\!\!\! \frac{x_{k}^+}{x_{k}^-} = e^{ip_k}.
\end{align}
The twists only manifest themselves as phase factors and upon sending them to zero we immediately reproduce the undeformed Bethe equations.

In order to compare the twisted Bethe equations for our models with \cite{Beisert:2005if,Beisert:2005he} we will rewrite them in terms of seven equations corresponding to the underlying $\alg{psu}(2,2|4)$ Dynkin diagram. However, as was shown in \cite{Martins:2007hb,de Leeuw:2007uf}, the equations given in \cite{Beisert:2005fw} agree with Bethe equations derived from string theory for $P=0$, while in general they can differ by factors of $e^{iP}$. For physical states in the untwisted theories that is not an issue, but here these factors have to be taken carefully into account. Let us follow the procedure outlined in \cite{Martins:2007hb} to bring the equations (\ref{eqn;FullBAE}-\ref{eqn;FullBAE3}) to the form presented in \cite{Beisert:2005fw}. Our equations are in the $\alg{sl}(2)$ grading and hence we will bring them to the form where $\eta = -1$. First, we identify the labels
\begin{align}
\label{eq:KKrelation}
&K^{\rm{I}} = K_4, && K^{\rm{II}} = K_1+K_3, && \dot{K}^{\rm{II}} = K_5+K_7, && K^{\rm{III}} = K_2,
&& \dot{K}^{\rm{III}} = K_6,
\end{align}
and we split the products accordingly, {\it i.e.}
\begin{align}
&\prod_{j=1}^{K^{\rm{II}}} \to \prod_{j=1}^{K_1}\ \ \prod_{j=1}^{K_3},
&&\prod_{j=1}^{\dot{K}^{\rm{II}}} \to \prod_{j=1}^{K_5}\ \ \prod_{j=1}^{K_7}.
\end{align}
Next we relabel the parameters in the following way
\begin{align}
&x^\pm_k = \frac{x_{4,k}}{g}, &&
y_k=\left\{\begin{array}{lc} x_{3,k}/g & \ k=1,\ldots K_3\\ g/x_{1,k} &  \ k=1,\ldots K_1 \end{array}\right.
&&
\dot{y}_k=\left\{\begin{array}{lc} x_{5,k}/g & \ k=1,\ldots K_5\\ g/x_{7,k} &  \ k=1,\ldots K_7 \end{array}\right.
\end{align}
Under this map, the main Bethe equation (\ref{eqn;FullBAE}) for fundamental particles becomes
\begin{align}
1=e^{-i(p_k L+\alpha+\dot{\alpha})}\prod_{j\neq k}^{K_4}S_0(p_j,p_k)
\prod_{j=1}^{K_3}\frac{x^+_{4,k}-x_{3,j}}{x^-_{4,k}-x_{3,j}}
\prod_{j=1}^{K_1}\!\frac{1 - \frac{g^2}{x^+_{4,k}x_{1,j}}}{1 - \frac{g^2}{x^-_{4,k}x_{1,j}}}
\prod_{j=1}^{K_5}\frac{x^+_{4,k}-x_{5,j}}{x^-_{4,k}-x_{5,j}}
\prod_{j=1}^{K_7}\!\frac{1 - \frac{g^2}{x^+_{4,k}x_{7,j}}}{1 - \frac{g^2}{x^-_{4,k}x_{7,j}}},
\end{align}
where $L=J-\frac{K_1-K_3-K_5+K_7}{2}$ and $S_0$ is the $\alg{sl}(2)$ scalar factor (see e.g. \cite{Arutyunov:2007tc}). To address the auxiliary Bethe equations we need to introduce the rapidities
\begin{align}
&w_k= u_{2,k}/g, && \dot{w}_k= u_{6,k}/g, && u_{i,j} \equiv x_{i,j} + \frac{g^2}{x_{i,j}}.
\end{align}
Then (\ref{eqn;FullBAE3}) is straightforwardly rewritten as (the dotted version is obtained by switching labels $K_{1,2,3}\to K_{7,6,5}$ and parameters similarly)
\begin{align}
1 = e^{2i\beta} \prod_{j\neq k}^{K_2}\frac{u_{2,k}-u_{2,j}+2i}{u_{2,k}-u_{2,j}-2i}
\prod_{j}^{K_1}\frac{u_{2,k}-u_{1,j}-i}{u_{2,k}-u_{1,j}+i}
\prod_{j}^{K_3}\frac{u_{2,k}-u_{1,j}-i}{u_{2,k}-u_{1,j}+i},
\end{align}
while (\ref{eqn;FullBAE2}) splits into two sets of equations for roots of type 1 and 3 (we have used the level matching condition $\prod_k\frac{x^+_k}{x^-_k} = e^{i\gamma}$)
\begin{align}
&1= e^{i(\alpha-\beta-\gamma/2)}\prod_{j=1}^{K_4}\frac{1-\frac{g^2}{x^-_{4,j}x_{1,k}}}{1-\frac{g^2}{x^+_{4,j}x_{1,k}}}
\prod_{j=1}^{K_2}\frac{u_{1,k}-u_{2,j}-i}{u_{1,k}-u_{2,j}+i}, &&
k = 1,\ldots K_1\\
&1= e^{i(\alpha-\beta+\gamma/2)}\prod_{j=1}^{K_4}\frac{x^-_{4,j}-x_{3,k}}{x^+_{4,j}-x_{3,k}}
\prod_{j=1}^{K_2}\frac{u_{3,k}-u_{2,j}-i}{u_{3,k}-u_{2,j}+i}, &&
k = 1,\ldots K_3,
\end{align}
again with similar equations for the dotted indices. We see that written this way our equations are exactly of the form of \cite{Beisert:2005fw} up to the twisting factors. These factors are precisely what was added to the Bethe equations in \cite{Beisert:2005if,Beisert:2005he}; next we will show explicit agreement with their twists for the various deformations.

\section{Explicit Models}
\label{sec:explicitmodels}

We are now ready to discuss specific twisted models and explicitly make a link to existing literature when applicable. We first focus on deformations of the sphere and then consider more general deformations that can involve all of $\ads$.

\subsection{Deformations of $S^5$}

\subsubsection*{Abelian orbifolds}

The twist corresponding to a generic abelian $\mathbb{Z}_S$ orbifold is represented by an element
\begin{align}
\mathrm{diag}(e^{-2\pi i T t_1/S}, e^{2\pi i( t_1-t_2)T/S}, e^{2\pi i( t_2-t_3)T/S}, e^{2\pi i t_3 T/S}),
\end{align}
where $T$ labels the twisted sector. From this we deduce that the angles transform as
\begin{align}
(\delta\phi_1,\delta\phi_2,\delta\phi_3) = \left(-\frac{2\pi t_2 T}{S},-\frac{2\pi(t_1-t_2+t_3)T}{S}, -\frac{2\pi(t_1-t_3)T}{S}\right),
\end{align}
resulting in the twist
\begin{align}\label{eqn;TwistOrbifold}
& \alpha=\dot{\alpha}=0, && \gamma = \frac{2\pi t_2 T}{S} \, , \nonumber\\
&\beta= \frac{\pi(2t_1-t_2)T}{S}, &&
\dot{\beta}= \frac{\pi(2t_3-t_2)T}{S}.
\end{align}
These twist are uniquely fixed by the geometry through the boundary conditions on the fields and in turn fix our Bethe equations (\ref{eqn;FullBAE}-\ref{eqn;FullBAE3}). These should then fully agree with \cite{Beisert:2005he} in the one-loop limit. Actually their result is readily generalized to all loop equations by supplementing \cite{Beisert:2005fw} with the appropriate phase factors.

Following \cite{Beisert:2005he} we should introduce phase factors $e^{2\pi i \frac{Ts_k}{S}}$ in front of the equations for roots of type $k$. In the $\alg{sl}(2)$ grading these twists are of the form \cite{Beccaria:2011qd}
\begin{align}
\nonumber
&s_1 =-t_1, && s_2 = 2t_1-t_2, && s_3 = t_2-t_1, &&s_4=0, && s_5 = t_2-t_3, && s_6 = 2t_3 - t_1, && s_7 = -t_3.
\end{align}
Our Bethe equations already contain phase factors, and comparing the above against our general twisted Bethe equations we find agreement exactly when
\begin{align}
& -\beta-\gamma/2 = 2\pi \frac{T s_1}{S}, && 2\beta = 2\pi \frac{T s_2}{S},
&& -\beta + \gamma/2 =  2\pi \frac{T s_3}{S},
\end{align}
with similar expressions for the dotted versions. Comparing this against (\ref{eqn;TwistOrbifold}) we see that they are in perfect agreement.

Since we are considering an orbifold, we should not forget to project onto the orbifold invariant part of the spectrum. This means we should consider states which satisfy \cite{Beisert:2005he}
\begin{align}
\label{eq:orbconstraintsphere}
t_2\, p + t_1\, q_2 +  t_3\, q_1 = 0\mod S.
\end{align}
Naturally this constraint can be phrased in more geometric terms via the relation of the $\alg{su}(4)$ weights $[q_1,p,q_2]$ to the angular momenta of the sphere \cite{arXiv:1103.2708}
\begin{align}
\label{eq:su4Js}
&J = J_1 = \frac{q_1+2p+q_2}{2},&& J_2= \frac{q_1+q_2}{2},&& J_3 = \frac{q_2-q_1}{2} \,,
\end{align}
so that \eqref{eq:orbconstraintsphere} becomes
\begin{equation}
\sum_{i=1}^3 J_i \, \delta \phi_i|_{T=1} = 0 \mod 2 \pi .
\end{equation}

\subsubsection*{$\gamma$-deformations} As discussed, TsT transformations of the sphere give rise to $\gamma$-deformed theories. Recalling that the twisted boundary conditions are parametrized by three deformation parameters $\gamma_{1,2,3}$ as
\begin{align}
\delta\phi_i=\phi_i(2\pi) - \phi_i(0) = -2\pi\epsilon_{ijk}\gamma_j J_k,
\end{align}
where the $J_i$ are the conserved Noether charges corresponding to the shift isometries in the $\phi_i$ direction on the sphere, these boundary conditions correspond to the twists
\begin{align}
& \alpha=\dot{\alpha}=0, && \gamma = 2\pi (\gamma_2 J_3 - \gamma_3 J_2) ,\nonumber\\
&\beta=\pi(\gamma_1 (J_2-J_3) +J_1(\gamma_3-\gamma _2)),&&
\dot{\beta}=\pi(J_1(\gamma_2 +\gamma_3) - \gamma_1 (J_2+J_3) ),
\end{align}
For $\gamma_1=\gamma_2=\gamma_3=\beta$ this reduces to the $\beta$-deformed theory for real $\beta$.

We will now compare the Bethe equations following from this twist to those derived in \cite{Beisert:2005if} or equivalently to \cite{Ahn:2011}. After a duality transformation to the $\alg{sl}(2)$ grading we find the following set of equations (see also the appendix of \cite{Ahn:2011} for the explicit $\alg{sl}(2)$ grading)
\begin{align}
{\scriptsize{
\begin{pmatrix}
-\gamma\\
\alpha-\beta-\gamma/2\\
2\beta\\
\alpha-\beta+\gamma/2\\
-\alpha-\dot{\alpha}\\
\dot{\alpha}-\dot{\beta}+\gamma/2\\
2\dot{\beta}\\
\dot{\alpha}-\dot{\beta}-\gamma/2
\end{pmatrix}=
\begin{pmatrix}
\delta_1 (K_5-2 K_6+K_7)-\delta_3 (K_1-2 K_2+K_3) \\
\delta_3 (K_0+K_2-K_3-K_5+K_6)-\delta_2 (K_5-2 K_6+K_7) \\
(\delta_1+2 \delta_2)(K_5-2 K_6+K_7) + \delta_3(2 K_5-2 K_6-2 K_0-K_1+K_3) \\
\delta_3(K_0+K_1-K_2-K_5+K_6)-(\delta_1 + \delta_2)(K_5-2 K_6+K_7) \\
0 \\
(\delta_2+\delta _3)(K_1-2 K_2+K_3)-\delta_1(K_0+K_2-K_3-K_6+K_7) \\
\delta_1(2 K_0+2 K_2-2 K_3-K_5+K_7)-(2\delta_2 + \delta_3)(K_1-2 K_2+K_3) \\
\delta_2(K_1-2 K_2+K_3) - \delta_1(K_0+K_2-K_3-K_5+K_6)
\end{pmatrix}
}},
\end{align}
where $K_i$ are the excitation labels from the Bethe equations and $K_0\equiv L$. The minus sign in the first term of the left hand side is due to the fact that $P = -(AK)_0$.

This over-determined system of equations indeed has a unique solution. In order to make contact with the twist discussed above we relate the labels to the conserved charges $J_i$ via \eqref{eq:su4Js}, and use their relation to the excitation numbers $K$ (see also \eqref{eq:KKrelation}) as
\begin{align}
&q_1 = \dot{K}^{\rm{II}} - 2 \dot{K}^{\rm{III}}, && q_2 = K^{\rm{II}} - 2 K^{\rm{III}}, && p=J-{\textstyle{\frac{1}{2}}}(K^{\rm{II}} + \dot{K}^{\rm{II}}) + K^{\rm{III}} + \dot{K}^{\rm{III}},\nonumber\\
&s_1 = K^{\rm{I}}-K^{\mathrm{II}}, &&s_2 = K^{\rm{I}}-\dot{K}^{\mathrm{II}}.
\end{align}
The parameters $\delta_i$ are related to the parameters $\gamma_i$ as
\begin{align}
&\delta_1 = \pi(\gamma_2 + \gamma_3)\,, &&\delta_2 = \pi(\gamma_1 -\gamma_2)\,, && \delta_3 = \pi(\gamma_2-\gamma_3)\, .
\end{align}
Putting this together correctly reproduces our twist.

\subsection{General deformations}

Let us now consider models involving the most generic set of boundary conditions.

\subsubsection*{General TsT transformations}

The twist for a general TsT transformation can be immediately read off from \eqref{eqn;genTsT} together with \eqref{gentwist1} and \eqref{gentwist2}. It is worthwhile to note that if the TsT transformation involves the time direction that the twist will become energy dependent. In other words the asymptotic Bethe equations pick up phases that depend on $E(p)$. This results in a very involved coupled system of equations.

\subsubsection*{General (abelian) orbifolds}

We will now discuss the generic orbifold $\mbox{AdS}_5/\Gamma^a \times \mbox{S}^5 /\Gamma^s$, with $\Gamma^a = \mathbb{Z}_R$ and $\Gamma^s = \mathbb{Z}_S$. Since we do not orbifold the $t$ direction $\gamma$ is unaffected by $\Gamma^a$ and consequently the twists $\beta,\dot{\beta}$ and $\gamma$ are again given by \eqref{eqn;TwistOrbifold}. We now quickly find the expressions for $\alpha,\dot{\alpha}$ by taking analogous expressions to those for $\beta,\dot{\beta}$ with the equivalent of $t_2$ put to zero. Labeling the twisted sectors for $\mathbb{Z}_R$ and $\mathbb{Z}_S$ by $\tilde{T}$ and $T$ respectively, this gives
\begin{align}
&\alpha= \frac{2\pi\, r_1 \tl{T}}{R}, && \dot{\alpha}= \frac{2\pi\, r_3 \tl{T}}{R} ,\nonumber\\
&\beta= \frac{\pi(2t_1-t_2)T}{S}, &&
\dot{\beta}= \frac{\pi(2t_3-t_2)T}{S},\nonumber\\
&\gamma = \frac{2\pi t_2 T}{S}. \label{eq:genorbtwist}
\end{align}
Orbifold invariance now requires
\begin{align}
t_2\, p + t_1\, q_2 + t_3\,q_1 + r_1\, s_2 + r_3\ s_1 & = 0\mod \mbox{LCM}(S,R)\,,\label{eq:orbconstraint}
\end{align}
where $\mbox{LCM}(S,R)$ denotes the least common multiple of $S$ and $R$ and $s_1$ and $s_2$ are the two spins of the conformal group. As mentioned earlier it is also entirely possible to describe non-abelian orbifolds in this framework. In each twisted sector we still get a twist of the form \eqref{eq:genorbtwist}. The only complication is in determining the physical states we should study, \textit{i.e.} in the analogue of \eqref{eq:orbconstraint}. We refer the interested reader to \cite{Solovyov:2007pw} where some specific examples are discussed in detail.

\section{The mirror TBA}
\label{sec:mirrorTBA}

\begin{figure}[h!]
\begin{center}
\includegraphics[width=12cm]{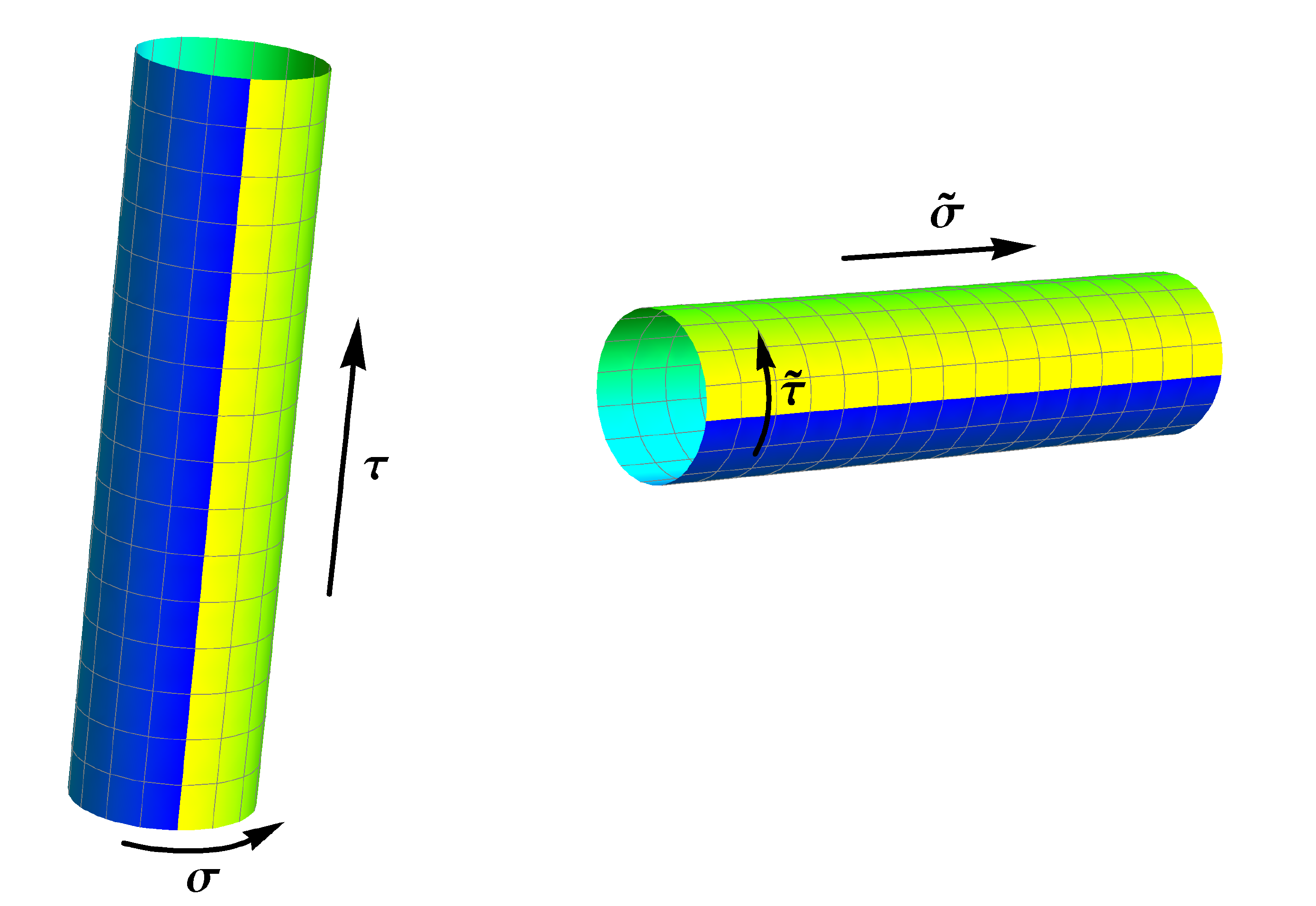}
\caption{A boundary condition in the spatial direction of string theory ($\sigma$) becomes a defect operator in the time direction ($\tilde{\tau}$) of the mirror model, here both illustrated by a color discontinuity.}
\label{fig:BCtoD}
\end{center}
\end{figure}

The deformations we are considering can be also readily embedded in the framework of the mirror Thermodynamic Bethe Ansatz (TBA), which is used to describe the spectrum of the undeformed superstring at finite size \cite{AF09b,GKKV09,BFT}. The TBA equations in principle contain non-trivial chemical potentials, but most of these are zero for the superstring with periodic boundary conditions\footnote{The chemical potential for the fermionic particles is $i \pi$, meaning we actually compute Witten's index and not quite the free energy for this theory.}. The effect of our deformations is to modify the boundary conditions on the string fields. These non-trivial boundary conditions can be viewed as a line defect running along the world-sheet cylinder of the superstring, and upon a double Wick rotation this induces a defect operator in the partition function of the mirror theory \cite{ITEP-89-144,Bajnok:2004jd} as illustrated in figure \ref{fig:BCtoD}.

Since our boundary conditions are in one to one correspondence with integrable twists, this defect operator commutes with the mirror $S$-matrix and simply leads to the appearance of chemical potentials in the TBA equations\footnote{Since the mirror scattering is unaffected this is the only change in the TBA equations.}. In other words the only modification to the mirror TBA equations \cite{AF09b,GKKV09,BFT} for the $\ads$ superstring is that we add the eigenvalue of the defect operator on a given particle to its equation. These considerations have already been applied to the $\gamma$-deformed theory in \cite{arXiv:1108.4914}, with the notable difference that there the authors approach the problem by means of a twisted $S$-matrix and a state independent boundary condition, rather than our twisted transfer matrix with a potentially state dependent boundary condition. We of course reproduce their TBA equations, just as the twisted S-matrix gives the same physical results as the twisted transfer matrix \cite{Ahn:2011}.

For the generic deformations introduced above we can immediately read off the eigenvalues of the twist element $g$ on the fundamental particles of the asymptotic Bethe ansatz\footnote{Note that this twist element $g$ should be taken for the charges of the state we wish to describe. In particular, for $\gamma$ deformations this means that the twist for the ground state only contains the angular momentum $J$, since all other charges vanish. In the twisted $S$-matrix approach this corresponds to the fact that the Drinfeld-Reshetikhin twist of the $\mbox{SU}(2|2)$ invariant $S$-matrix does not contribute to the ground state energy, leaving the state independent boundary condition which depends on $J$ only.}. The eigenvalue of $g$ acting on a string complex or bound state is of course simply a sum of the eigenvalues on its constituents. For the reader's convenience we have summarized these eigenvalues\footnote{The eigenvalue for $Q$-particles is actually the negative of the chemical potential presented in table \ref{tab:chempot}; with this definition the chemical potentials enter uniformly in the TBA equations of \cite{AF09b} that will follow shortly.} in table \ref{tab:chempot}; the physical chemical potentials are $\mu_{1|w^\pm}$ and $\mu_{1|vw^\pm}$.

\begin{table}[h!]
\begin{center}
\begin{tabular}{|c|c|}
\hline  $\chi$ & $\mu_\chi$ \\
\hline $M|w^+$   & $2 i M \beta$    \\
\hline $M|vw^+$   & $2 i M \alpha$    \\
\hline $y^+$   & $i(\alpha-\beta)$ \\
\hline $M|w^-$   & $2 i M \dot{\beta}$  \\
\hline $M|vw^-$    & $2 i M \dot{\alpha}$  \\
\hline $y^-$    & $i(\dot{\alpha}-\dot{\beta})$ \\
\hline $Q$   & $i Q (\alpha+\dot{\alpha})$   \\
\hline
\end{tabular}
\end{center}
\caption{Chemical potentials for particles of type $\chi$ of the mirror TBA. Note that for TsT deformations $\alpha$ and $\beta$ depend on the state under consideration.}
\label{tab:chempot}
\end{table}

\noindent These eigenvalues are defined modulo $2 \pi i$ because they are read off from the eigenvalues $e^{\mu_\chi}$ of the defect operator. The chemical potentials themselves are unambiguously defined because of their direct link to the quasi-periodic boundary conditions for fundamental particles in the string theory.

Before moving on to the TBA equations, we should note that the partition function of the mirror theory with purely imaginary chemical potentials is not well defined. In fact, the twists $\alpha$ and $\beta$ should have a small positive imaginary part in order to suppress large $M$ magnonic contributions to the partition function. With this regularization the theory is well defined, and we can consider the canonical ground state TBA equations. The resulting considerations are somewhat subtle, but the upshot is a set of simplified and hybrid TBA equations together with a set of large $u$ asymptotics for the Y-functions that are well defined when the regulator is taken away and can hence be taken as proper tools to study the spectral problem.

\subsection{The TBA equations}

The canonical ground state TBA equations with non-zero chemical potentials are given by
\begin{align}
\log Y^{(a)}_{M|w} =& - \mu^{(a)}_{M|w} + \log\big(1+\frac{1}{Y^{(a)}_{N|w}}\big)\star K_{NM} + \log \frac{1-\frac{1}{Y^{(a)}_-}}{1-\frac{1}{Y^{(a)}_+}} \hat{\star} K_M \, ,\\
\log Y^{(a)}_{M|vw} =& - \mu^{(a)}_{M|vw} + \log\big( 1+\frac{1}{Y^{(a)}_{N|vw}}\big)\star K_{NM} + \log \frac{1-\frac{1}{Y^{(a)}_-}}{1-\frac{1}{Y^{(a)}_+}}\hat{\star}  K_M \nonumber \\ & \quad - \log (1+Y_Q)\star  K^{QM}_{xv}  \, ,\\
\log {Y^{(a)}_\pm} =& - \mu^{(a)}_\pm  - \log\big(1+Y_Q \big)\star  K^{Qy}_\pm + \log \frac{1+\frac{1}{Y^{(a)}_{M|vw}}}{1+\frac{1}{Y^{(a)}_{M|w}}}\star  K_M\,,\\
\log Y_Q =& \, - \mu_Q - J \, \tilde{\mathcal{E}}_{Q} + \log\big(1+Y_{M} \big) \star K_{\mathfrak{sl}(2)}^{MQ}
\\[1mm]
&\quad + \sum_{(a)=\pm} \log\big(1+ \frac{1}{Y^{(a)}_{M|vw}} \big) \star  K^{MQ}_{vwx} + \log \big(1- \frac{1}{Y^{(a)}_\pm}\big) \hat{\star}  K^{yQ}_\pm   \, , \nonumber
\end{align}
where the $\mu_\chi$ are given in table \ref{tab:chempot} keeping in mind the implicit regulator. For a complete list of the kernels and a derivation of these equations without chemical potentials we refer the reader to \cite{AF09b}. The ground state energy is given by
\begin{equation}
\label{eq:gse}
E=-\frac{1}{2\pi}\int\, du\, \frac{d\tilde{p}_Q}{du}\log(1+Y_Q)\, .
\end{equation}

As we will see below, it is useful to transform the canonical TBA equations to an alternate form. In particular the equations for $w$ and $vw$ strings can be simplified\footnote{The equation for $Q$ particles can be simplified as well, but the hybrid form we present below is more suitable for various computations.} by application of the kernel $(K+1)^{-1}$\cite{AF09b,Arutyunov:2009ux}. The chemical potentials are in the kernel of this operator, and hence the simplified equations do not depend explicitly on them. These simplified equations are given by
\begin{align}
\log Y_{M|w}^{{(a)}}  = & \, I_{MN}\log(1+Y_{N|w}^{{(a)}})\star s+
\delta_{M1}\log\frac{1-\frac{1}{Y_{-}^{{(a)}}}}{1-
\frac{1}{Y_{+}^{{(a)}}}}\hat{\star}s \,,\label{eq:sTBAwbefhyb}\\
\log Y_{M|vw}^{{(a)}}  = & \,-\log(1+Y_{M+1})\star s+I_{MN}\log(1+Y_{N|vw}^{
{(a)}})\star s+\delta_{M1}\log\frac{1-Y_{-}^{{(a)}}}{1-Y_{+}^{{(a)}}}
\hat{\star}s\,.\label{eq:sTBAvwbefhyb}
\end{align}
These equations together with the canonical TBA equations allow us to find the asymptotic behavior of the $Y_{M|w}$ and $Y_{M|vw}$ functions as we will show shortly. Then, with this asymptotic behavior and the simplified equations, we can eliminate the infinite sums over $w$ and $vw$ strings from the equations for $y$ and $Q$ particles to bring them to their hybrid form \cite{AFS09}. The only modification in the derivation of the hybrid TBA equations is in the presence of chemical potentials and we discuss the resulting contribution in \ref{app:hybridTBA}. The upshot is the following set of equations
\begin{align}
\log Y_{M|w}^{{(a)}}  = & \, I_{MN}\log(1+Y_{N|w}^{{(a)}})\star s+
\delta_{M1}\log\frac{1-\frac{1}{Y_{-}^{{(a)}}}}{1-
\frac{1}{Y_{+}^{{(a)}}}}\hat{\star}s \,,
\label{eq:sTBAw}\\
\log Y_{M|vw}^{{(a)}}  = & \,-\log(1+Y_{M+1})\star s+I_{MN}\log(1+Y_{N|vw}^{
{(a)}})\star s+\delta_{M1}\log\frac{1-Y_{-}^{{(a)}}}{1-Y_{+}^{{(a)}}}
\hat{\star}s\,,  \label{eq:sTBAvw}\\
\log\frac{Y_{+}^{{(a)}}}{Y_{-}^{{(a)}}} = &
\log(1+Y_{Q})\star K_{Qy}\,, \label{eq:sTBAyd} \\
\log Y_{-}^{{(a)}}Y_{+}^{{(a)}}  = & \,-\log(1+Y_{Q})\star K_{Q}+
2\log(1+Y_{Q})\star K_{xv}^{Q1}\star s+2\log\frac{1+Y_{1|vw}^{
{(a)}}}{1+Y_{1|w}^{{(a)}}}\star s\,,  \label{eq:sTBAyp}\\
\log Y_Q  = & \,-L\tilde{\mathcal{E}}_Q+\log(1+Y_{Q^\prime})\star
\left(K_{sl(2)}^{Q^\prime Q}+2s\star K_{vx}^{Q^\prime -1,Q}\right)\nonumber \\
  & +\sum_{{(a)}=\pm}\biggr[\log\biggl(1+Y_{1|vw}^{{(a)}}\biggr)
\star s\,\hat{\star}K_{yQ}+\log(1+Y_{Q-1|vw}^{{(a)}})\star s \nonumber \\
& \hspace{50pt} -\log
\frac{1-Y_{-}^{{(a)}}}{1-Y_{+}^{{(a)}}}\hat{\star}s\star K_{vwx}^{1Q}
 +\frac{1}{2}\log
\frac{1-\frac{1}{Y_{-}^{{(a)}}}}{1-\frac{1}{Y_{+}^{{(a)}}}}
\hat{\star}K_Q
\nonumber\\
& \hspace{70pt} +\frac{1}{2}\log(1-\frac{1}{Y_{-}^{{(a)}}})
(1-\frac{1}{Y_{+}^{{(a)}}})\hat{\star}K_{yQ}\biggl]\,, \label{eq:sTBAQ}
\end{align}
From these equations and the large $u$ asymptotics for the Y-functions for $w$ and $vw$ strings we can directly read off the large $u$ asymptotics of all Y-functions\footnote{They can of course also be derived from the canonical equations.}. We see that the chemical potentials have disappeared from this form of the TBA equations, meaning they are unchanged from the original model as already argued in \cite{deLeeuw:2011rw}. Consequently also the $Y$-system is unchanged for these deformations. This also follows from the fact that the chemical potentials satisfy the conditions to obtain an undeformed $Y$-system \cite{arXiv:0901.3753}, formulated in \cite{Cavaglia:2010nm,Cavaglia:2011kd}. We would like to emphasize again that while the chemical potentials disappear from the simplified TBA and $Y$-system equations, the Y-functions must satisfy the \emph{canonical} TBA equations, which translates to a set of large $u$ asymptotics on them.

\subsection{The ground state solution}
\label{sec:GS}

The first concrete state we need to consider in a generically deformed theory is the ground state. The Y-functions for this state will give us the concrete large $u$ asymptotics of the Y-functions for any excited state with the same chemical potentials. We start by considering the TBA equations in the asymptotic limit. In this limit the $Y_Q$ functions are exponentially small, and from the canonical TBA equations it follows that $Y_+ = Y_-$ since their chemical potentials are equal. Then then simplified TBA equations for $w$ and $vw$ strings (\ref{eq:sTBAwbefhyb},\ref{eq:sTBAvwbefhyb}), become simple recursion relations for constant Y-functions
\begin{equation}
Y^{\circ}_{M+1|w} = \frac{Y^{\circ 2}_{M|w}}{1+Y^{\circ}_{M-1|w}} -1 \, ,
\end{equation}
with an identical equation for $vw$ strings, and we note that $Y^{\circ}_{0|(v)w}$ is zero by definition. By this recursion relation all $Y^{\circ}_{M|w}$ are uniquely fixed in terms of the value of $Y^{\circ}_{1|w}$. The constant solution of this equation which also satisfies the canonical TBA equations with zero chemical potentials is given by \cite{arXiv:0906.0499}
\begin{equation}
Y^{\circ}_{M|w} = M(M+2) \, ,
\end{equation}
which satisfies the recursion relation for the simple reason that $M^2 = (M+1)(M-1)+1$. A clear generalization of this solution is
\begin{equation}
Y^{\circ+}_{M|w} = [M]_{q}[M+2]_{q} \, ,
\end{equation}
where we have introduced $q$-numbers $[n]_q$ as
\begin{equation}
[n]_q = \frac{q^n - q^{-n}}{q-q^{-1}} \, ,
\end{equation}
since $q$-numbers retain the property $[M]_q^2 = [M+1]_q[M-1]_q+1$ for any $q \in \mathbb{C}$. By picking $q$ appropriately, we can obtain any desired constant value for $Y^{\circ}_{1|w}$, and hence this is the general constant solution of the simplified TBA equations. What remains is to determine the value of $q$ such that these Y-functions also satisfy their canonical TBA equations. We do this in \ref{app:hybridTBA}, and by substituting the result in the hybrid equations (\ref{eq:sTBAyd},\ref{eq:sTBAyp},\ref{eq:sTBAQ}) we find the following asymptotic auxiliary Y-functions
\begin{align}
Y^{\circ+}_{M|w} & = [M]_{q_{\beta}}[M+2]_{q_{\beta}} \,,&   Y^{\circ-}_{M|w} & = [M]_{q_{\dot{\beta}}}[M+2]_{q_{\dot{\beta}}} \, , \label{eq:YwGSasympt}\\
Y^{\circ+}_{M|vw}&  = [M]_{q_{\alpha}}[M+2]_{q_{\alpha}} \,,& Y^{\circ-}_{M|vw} & = [M]_{q_{\dot{\alpha}}}[M+2]_{q_{\dot{\alpha}}} \, , \label{eq:YvwGSasympt}\\
Y^{\circ+}_{\pm}&  = [2]_{q_{\alpha}}/[2]_{q_{\beta}} \,,& Y^{\circ-}_{\pm} & = [2]_{q_{\dot{\alpha}}}/[2]_{q_{\dot{\beta}}} \, ,
\label{eq:YpmGSasympt}
\end{align}
while the main Y-functions are
\begin{equation}
\label{eq:YQasympt}
Y^\circ_{Q} =  ([2]_{q_{\alpha}}-[2]_{q_{\beta}})([2]_{q_{\dot{\alpha}}}-[2]_{q_{\dot{\beta}}})[Q]_{q_{\alpha}}[Q]_{q_{\dot{\alpha}}} e^{-J \tilde{\mathcal{E}}_Q(\tilde{p})}\, ,
\end{equation}
where the $q$ have become phases denoted $q_\theta \equiv e^{i \theta}$. This asymptotic solution is a generalization of the asymptotic solution presented in \cite{deLeeuw:2011rw,arXiv:1108.4914} for deformations of the sphere. Of course, this asymptotic solution also follows directly from the twisted transfer matrix. For example, the asymptotic $Y_Q$-function is directly given by
\begin{align}
Y^\circ_Q = \chi(\pi(G))\chi(\pi(\dot{G}))e^{-J\tilde{\mathcal{E}}_Q},
\end{align}
where $\chi(\pi(G)) = \mathrm{tr}(\pi(G))$ is the character of the twist element $G$ in the $Q$-particle bound state representation in the transfer matrix (\ref{eq:twistedtfgeneral}); the twisted transfer matrix for zero particles. Naturally this agrees with (\ref{eq:YQasympt}). Just as for the chemical potential, here we should keep in mind that the charges entering the twisted transfer matrix should be those of the ground state.

\subsection{Large $u$ asymptotics and excited states}

In the large $u$ limit, the convolutions on the right hand side of the canonical TBA equations become insensitive to fluctuations of the Y-functions around the origin, and only their constant `background' values play a role. It follows that for any state with a given set of chemical potentials, these constant values should be given by the asymptotic ground state solution. In other words
\begin{equation}
\label{eq:uasymp}
\lim_{u\rightarrow \pm \infty} Y_\chi(u) = Y^\circ_\chi \, ,
\end{equation}
where $Y^\circ_\chi$ denotes the asymptotic ground state values of the Y-function of type $\chi$, as given in equations (\ref{eq:YwGSasympt}-\ref{eq:YQasympt}). Note that while the regulator plays an essential role in the canonical TBA equations, the regulator can be taken away by describing the spectral problem through the simplified TBA equations and the asymptotics (\ref{eq:uasymp}).

Coming to excited states, as we saw above the twisted transfer matrix can be used to construct the asymptotic Y-functions for the ground state. The same is true for excited states and allows us to apply the contour deformation trick \cite{AFS09,arXiv:1103.2708} to obtain excited state TBA equations from the ground state equations. The differences in analytic structure that distinguish the different excited states, i.e. the number and location of poles and zeroes of the Y-functions, can be determined from the asymptotic Y-functions constructed through the twisted transfer matrix. This idea has already been used to find the excited state TBA equations for the $\alg{sl}(2)$ descendant of the Konishi operator upon orbifolding the sphere \cite{deLeeuw:2011rw}. We would like to point out that for orbifolds the asymptotics (\ref{eq:uasymp}) are the same for any state in the same model, but this is not the case for theories obtained by TsT transformations as the chemical potentials depend on the state under consideration.

The simplified ground state TBA equations naturally have real Y-functions as solutions, and indeed the asymptotic solution (\ref{eq:YwGSasympt}-\ref{eq:YQasympt}) is real. Reality of the asymptotic Y-functions for excited states follows by the way the twist parameters enter the twisted transfer matrix \eqref{eqn;FullEignvalue}; the terms in the undeformed transfer matrix that are related by conjugation are now multiplied by conjugate phases. The exact Y-functions that solve the excited state TBA equations are therefore also real, since the equations are obtained by the contour deformation trick with a real asymptotic solution.

\section{Wrapping corrections}

In this section we discuss leading order wrapping corrections to the ground state energy for generic deformations. We also compute wrapping corrections to the energy of a few two particle states an orbifold of $\mbox{AdS}_5$.

\subsection{The deformed ground state}

With the asymptotic ground state solution solution we can perturbatively solve the TBA equations and find the energy of the ground state. From the structure of the $Y^\circ_Q$ functions we see that their expansion starts at order $g^{2J}$, known as single wrapping order. Next, expanding the TBA equations it is easy to see that the first correction to the asymptotic form of the $Y_Q$-functions comes in at order $g^{4J}$ or double wrapping order. Hence up to this order the perturbative ground state energy is found by simply expanding (\ref{eq:gse}) to the desired order in $g$.

Let us first give the leading order wrapping correction
\begin{align}
\label{eq:GSsinglewrapping}
E =& \frac{(\cos \alpha-\cos \beta)(\cos \dot{\alpha}-\cos \dot{\beta})}{\sin\alpha\sin\dot{\alpha}}
\frac{\Gamma(J-\frac{1}{2})}{2\sqrt{\pi}\Gamma(J)}\times \\
&\times\left[
\mathrm{Li}_{2J-1} (e^{i(\alpha+\dot{\alpha})}) +
\mathrm{Li}_{2J-1} (e^{-i(\alpha+\dot{\alpha})}) -
\mathrm{Li}_{2J-1} (e^{i(\alpha-\dot{\alpha})}) -
\mathrm{Li}_{2J-1} (e^{-i(\alpha-\dot{\alpha})})\right]\nonumber \, .
\end{align}
Here and below $\mathrm{Li}_{n}$ is the polylogarithm. For $\alpha = \dot{\alpha} =0$ this indeed reduces to the energy of the orbifold ground state \cite{deLeeuw:2011rw}. This formula can actually be extended up to order $g^{4J}$ which is the double wrapping order.  It is instructive to write the asymptotic Y-function explicitly as function of the mirror momentum $\tilde{p}$:
\begin{align}\label{eqn;AsymYQ}
Y^{\circ}_{Q}(\tl{p}) = \chi(G)\chi(\dot{G}) \left[ \frac{2 g}{\sqrt{Q^2+\tl{p}^2}+\sqrt{4g^2 +Q^2+\tl{p}^2}}\right]^{2J}.
\end{align}
To order $g^{4J}$  (\ref{eq:gse}) simply reduces to
\begin{align}
E = \sum_{Q=1}^{\infty}\int d\tl{p}\, Y^{\circ}_{Q}(\tl{p}).
\end{align}
Evaluating the integral perturbatively in $g$ we find
\begin{align}
\label{eq:GSfullsinglewrap}
E =- \sum_{m=1}^{\infty}\frac{J (-1)^m (2g)^{2(J + m)}}{4}
\frac{ \Gamma (J+m-\frac{3}{2}) \Gamma (J+m-\frac{1}{2})}{\Gamma (m) \Gamma (2 J+m)} \sum_{Q=1}^{\infty}\frac{\chi(G)\chi(\dot{G})}{Q^{2 (J+ m)-3} }.
\end{align}
The sum over $Q$ yields a term similar to \eqref{eq:GSsinglewrapping}
\begin{align}
\sum_{Q=1}^{\infty}\frac{\chi(G)\chi(\dot{G})}{Q^{x} } =& \frac{(\cos \alpha-\cos \beta)(\cos \dot{\alpha}-\cos \dot{\beta})}{\sin\alpha\sin\dot{\alpha}} \times\\
&\times\left[
\mathrm{Li}_{x} (e^{i(\alpha+\dot{\alpha})}) +
\mathrm{Li}_{x} (e^{-i(\alpha+\dot{\alpha})}) -
\mathrm{Li}_{x} (e^{i(\alpha-\dot{\alpha})}) -
\mathrm{Li}_{x} (e^{-i(\alpha-\dot{\alpha})})\right]\nonumber.
\end{align}
In the limit of undeformed anti-de Sitter space, (\ref{eq:GSfullsinglewrap}) reduces to the single wrapping correction in equation (5.5) of \cite{arXiv:1108.4914}. Going beyond this order requires a significant amount of computation due to the non-trivial corrections to the Y-functions. Nonetheless, the full double wrapping L\"uscher-type correction for the $\gamma$-deformed ground state energy was recently compared to the corresponding TBA contribution, and shown to be in full agreement \cite{arXiv:1108.4914}. The relevant computations generalize partially to more generic deformations; their results also apply to orbifolds of the sphere. Moreover, in \cite{arXiv:1108.4914} the lowest order double wrapping correction to the length three ground state was explicitly found to be
\begin{align}
E_{0}(3)  = & -(2-[2]_{q_{\beta}})(2-[2]_{q_{\dot{\beta}}})\left(\frac{3}{16}
\zeta_{3}g^{6}-\frac{15}{16}\zeta_{5}g^{8}+\frac{945}{256}
\zeta_{7}g^{10}-\frac{3465}{256}\zeta_{9}g^{12}+\dots\right)\nonumber \\
   & -(2-[2]_{q_{\beta}})(2-[2]_{q_{\dot{\beta}}})\left([2]_{q_{\beta}}+[2]_{q_{\dot{\beta}}}-4)
\right)\frac{15}{256}\zeta_{3}\zeta_{5}g^{12} + \dots \nonumber\\
   & +(2-[2]_{q_{\beta}})^{2}(2-[2]_{q_{\dot{\beta}}})^{2}\left(-
\frac{9}{256}\zeta_{3}^{2}+
\frac{189}{4096}\zeta_{7}\right)g^{12}+\dots \, .
\end{align}
where we have trivially generalized their result to orbifolds of the sphere in addition to $\gamma$-deformations, parametrizing the deformation by $\beta$ and $\dot{\beta}$. To generalize this result to deformations of the anti-de Sitter space would likely require considerable computational effort, as already the single wrapping correction (\ref{eq:GSsinglewrapping}) is significantly more involved\footnote{In the final details of their computation, the authors of  \cite{arXiv:1108.4914} found a single term through an accurate numerical computation which was then expressed in a transcendental basis using the program EZ-face (documented in \cite{math/9910045}). Given the form of the single wrapping correction for more generic deformations we would likely encounter numbers not even expressible in terms of multiple zeta values, in fact this `number' would be a function of $\alpha$ and $\dot{\alpha}$.}. So far we have focussed on completely generic deformations, however in special cases there are certain interesting observations to be made.

\subsection{Observations on the ground state solution}
\label{sec:GSobservations}

Here we would like to make a few comments based on the asymptotic ground state solution (\ref{eq:YwGSasympt}-\ref{eq:YQasympt}) and the corresponding energy corrections (\ref{eq:GSsinglewrapping}). Let us start with a small but relevant point.

\subsubsection*{Regularizing the ground state energy}

In \cite{arXiv:0906.0499} the authors discussed the vanishing of the ground state energy of the undeformed theory at any finite size. However they observed a curious divergence of the ground state energy at $J=2$. This divergence was subsequently also found in the case of orbifolds \cite{deLeeuw:2011rw} and $\gamma$-deformations \cite{arXiv:1108.4914}.

In order to show vanishing of the ground state energy, we would basically like to show that the asymptotic solution (here for zero twist) is exact, \textit{i.e.} that vanishing $Y_Q$ functions are a solution of the TBA equations. In order to show this the authors of \cite{arXiv:0906.0499} introduced a regularization in the form of a small chemical potential, denoted $h$, which made the $Y_Q$ functions $O(h^2)$ such that they vanish in the $h \rightarrow 0$ limit. However for $J=2$ this limit does not commute with the infinite sum in the energy formula leading to divergent energy even though the $Y_Q$ functions vanish. Since the chemical potential $h$ directly corresponds to a twist $\beta =\dot{\beta} =h$, this divergence is indeed reproduced by our energy formula (\ref{eq:GSsinglewrapping}) which diverges as $\alpha,\dot{\alpha} \rightarrow 0$ for non-zero $\beta$ and $\dot{\beta}$. However (\ref{eq:GSsinglewrapping}) does not diverge as $\alpha,\dot{\alpha} \rightarrow 0$ for $\beta = \dot{\beta} = 0$, meaning that a chemical potential in the $\mbox{AdS}$ sector regularizes the ground state solution for the undeformed theory, showing that the ground state energy vanishes without any divergence as it should.

While this provides us with a way to regularizes the ground state energy of the undeformed theory, the divergence of the ground state energy at $J=2$ for certain orbifolds and $\gamma$-deformed theories remains an open question.

\subsubsection*{Residual supersymmetry}

As we saw above, the asymptotic ground state solution is an exact solution of the TBA equations when the corresponding $Y_Q$ functions vanish in some appropriately regularized sense. In addition to the undeformed superstring, this situation arises when we deform parts of anti-de Sitter space and the sphere in an identical fashion. By this we mean a twist where at least two group elements are identical, \textit{i.e.} $\alpha = \pm \beta$ or $\dot{\alpha} = \pm \dot{\beta}$. Indeed, in this limit the asymptotic $Y^\circ_Q$ functions (\ref{eq:YQasympt}) and correspondingly the ground state energy (\ref{eq:GSsinglewrapping}) vanish. In other words, just as for the undeformed ground state, the corresponding asymptotic solution is an exact solution of the TBA equations. Naturally this is not an accident but owes to the residual symmetry left in this situation.

In the undeformed theory the ground state is a half-BPS state, meaning it is annihilated by a set of supercharges
\begin{equation}
Q^{1,2}_\gamma | 0 \rangle  = \bar{Q}^{3,4}_{\dot{\gamma}} | 0 \rangle  = 0 \, , \, \, \, \gamma = 1,2.
\end{equation}
Combined with the fact that it is a highest weight state this allows us to derive a relation between the scaling dimension and other weights of the state from the superconformal algebra in the usual fashion, in particular by considering the anti-commutator $\{ Q, S \}$ \cite{hep-th/0209056}
\begin{equation}
\label{eq:bpscomm}
\{ Q^i_\gamma, S_j^\delta \} = 4 \delta_i^j (M_\gamma^\delta -\tfrac{i}{2} D) - 4 \delta_\gamma^\delta R^i_j
\end{equation}
where the $M_\gamma^\delta$ are the generators of the conformal group corresponding to the $\mbox{SU}(2)$ subgroup with spin $s_1$ and the $R^i_j$ are the generators of $\mbox{SU}(4)$ and $D$ is the dilatation operator. Indeed, since the left hand side annihilates a highest weight half-BPS state for $i=1,2$ and the action of the $R^i_i$ is expressible via the corresponding weights, we get a relation between the scaling dimension and the $\alg{su}(4)$ weights $[q_1,p,q_2]$. As a result the state cannot have an anomalous dimension; its scaling dimension is protected.

In a deformed theory, the ground state can still be considered half-BPS in the sense that it is annihilated by certain operators, but those operators have lost their meaning and relation to the scaling dimension since the superconformal algebra is broken. This means there are generically no protected operators. However, precisely deformations of the type $\alpha = \pm \beta$ or $\dot{\alpha} = \pm \dot{\beta}$ preserve part of the superconformal algebra. In particular some of the relations (\ref{eq:bpscomm}) which protect the scaling dimension of the ground state remain. Concretely we preserve the relations (\ref{eq:bpscomm}) for $i=j=1,2$, $\gamma=\delta=2,1,$ if we twist with $\alpha=\beta$ and for $i=j=\gamma=\delta=1,2,$ if we twist with $\alpha=-\beta$. The relations are similar for the $\bar{Q}$s and $\bar{S}$s with dotted twists. As a result the ground state energy is still a protected quantity in such deformed models.

\subsubsection*{Protected states without supersymmetry}

In addition to the above, there is another class of deformed models with a ground state energy which does not receive corrections. The most interesting representative of this class is $\gamma$-deformed theory with $\gamma_2=\gamma_3=0$. This feature arises because the ground state solution for generic $\gamma$-deformed theories is independent of $\gamma_1$ as immediately follows from our twisted transfer matrix since the ground state has only charge $J=J_1$. This matches with the results of \cite{arXiv:1108.4914} since in the twisted $S$-matrix approach the ground state energy is independent of the Drinfeld-Reshetikhin twist, which in this case carries all dependence on $\gamma_1$. Since the $Y_Q$ functions and ground state energy vanish in the limit $\gamma_{2,3} \rightarrow 0$, we have an exact solution of the TBA equations with zero energy. However, except for the cases $\{\gamma_1,\gamma_2,\gamma_3\} = \{\gamma,\pm \gamma,\pm \gamma\}$ (with all possible choices of signs) $\gamma$-deformations break all supersymmetry. Therefore we have no immediate explanation for the apparent `protection' of the ground state energy, and it would be interesting to understand how this arises from the point of view of the deformed theory\footnote{By mapping it to the undeformed theory which has periodic boundary conditions for the ground state, this result is of course immediate.}. Other models of this type would be generic TsT transformed backgrounds, with the parameters that couple to $J$ put to zero.

\subsection{Two particle states and AdS orbifolds}
\label{sec:adsorb}

As we have seen above, deformations of anti-de Sitter space can be useful, but also seem to rather directly increase the technical complexity of the resulting computations. As an example of this, wrapping corrections become quite involved for generic $\mbox{SU}(2)\times\mbox{SU}(2)$ orbifolds of $\mbox{AdS}_5$, while for the sphere they can be readily computed for reasonable states \cite{arXiv:1009.4118,deLeeuw:2011rw}. Here we would like to briefly illustrate the different features wrapping corrections for orbifolds of $\mbox{AdS}_5$ have and how quickly the expressions become rather unpleasant.

Since we do not want to deform the sphere or orbifold in time-like directions we have $\gamma=0$, {\it i.e.} the level-matching condition is unchanged and physical states are still required to have vanishing total momentum. In the $\alg{sl}(2)$ sector we see from the main Bethe equations that this is compatible with level-matching only if $\alpha=-\dot{\alpha}$. We restrict to this choice and set
\begin{align}
&\alpha = -\dot{\alpha} = 2\pi \frac{r}{R}, && r,R \in\mathbb{N}.
\end{align}
The first interesting state to consider is of course the ground state. Its energy is non-zero due to wrapping effects and is readily obtained by plugging in the explicit twist into \eqref{eq:GSsinglewrapping}. One particle states in this sector have the same energy as the ground state as a consequence of level matching.

Next we arrive at two-particle states. Due to our choice of twists, the Bethe equations coincide with the undeformed ones which to lowest order in $g$ are solved by the usual rapidities
\begin{align}\label{eqn;rapsl2}
u_1 = -u_2 = \cot \frac{n\pi}{J+1},
\end{align}
where $n$ runs up to $\lfloor \frac{J}{2} \rfloor$. The general transfer matrix for these states is rather involved, although it is obtained in a straightforward fashion, and we will not present it explicitly. We computed the resulting wrapping corrections for the first few values of $J$ for a $\mathbb{Z}_4$ orbifold and have presented them for $J=2,3$ in table \ref{tab:wrapcorrs}.
\begin{table}
\begin{tabular}{|c||c|c|c|c|}
\hline
$J$ &  $\alpha=\pi$ & $\alpha=\pi/2$ & $\alpha=\pi/4$\\
\hline
$2$ & $-\frac{1}{9}$ & $\frac{6-7\zeta(3)}{32}$ & $\frac{(3-2\sqrt{2})(48 K+18\ln 2-15-35\zeta(3))}{16}$\\
\hline
$3$ & $-\frac{1+6\zeta(3)}{18}$ & $\frac{1-4\ln 2}{16}$ & $\frac{(3-2\sqrt{2})(2048K+512\ln 2-256-288\zeta(3)+1581\zeta(5) -2\psi^{(3)}(\frac{1}{4}) + 2\psi^{(3)}(\frac{3}{4})}{4096}$\\
\hline
\end{tabular}
\label{tab:wrapcorrs}
\caption{The values of $E_{LO}/g^{2J}$ for $J=2,3$ for our $\mathbb{Z}_4$ orbifold of $\mbox{AdS}_5$. $K$ is the Catalan number.}
\end{table}
We find that the complexity of the finite size corrections increase rather rapidly when taking more general values of $\alpha$ and we were not able to derive a closed expression for generic $\alpha$, which is readily possible for the analogous deformation involving $\beta$. We also considered simultaneously orbifolding the sphere such that we partially restore supersymmetry, and as expected the wrapping corrections show up at a higher order. Unfortunately the resulting expressions are rather unsightly and we choose not to present them here. It would be interesting to extend this discussion to generic twist two and three operators, along the lines of \cite{deLeeuw:2010ed,deLeeuw:2011rw,Beccaria:2011qd} where this was considered for deformations of the sphere.

\section{Conclusion}
\label{sec:conclusion}

In this paper we have presented a unified treatment of the finite size integrability of strings on orbifolds and TsT transformed versions of $\ads$.  Firstly, through the twisted transfer matrix \eqref{eqn;FullEignvalue} we provide the necessary ingredients for the computation of wrapping corrections for arbitrary operators in the dual deformed gauge theories. Moreover, with the twisted asymptotic solution the TBA equations provide a means to compute the spectrum at truly finite size in these theories, along the lines of e.g. \cite {Gromov:2009zb,Frolov:2010wt}. In this paper we illustrated our discussion by a computation of the leading order wrapping correction to the ground state energy for generic deformations and to the energy of certain two particle states in the $\alg{sl}(2)$ for a $\mathbb{Z}_4$ orbifold of $\mbox{AdS}_5$. Two open interesting questions that explicitly arise have to do with the apparent protection of the ground state in certain non-supersymmetric theories and the divergence of this energy at length two for deformations of the sphere. We have also shown how deformations of $\mbox{AdS}_5$ can provide a useful extra parameter in the AdS/CFT spectral problem, for example by regularizing the ground state energy. Furthermore, as we indicated in the main text, there is a mismatch with a few results in the literature which should be understood. Of course it would also be very interesting to see a more proper investigation of the AdS/CFT duals for deformations of $\mbox{AdS}_5$ since the integrability properties from the string theory point of view, though technically more involved, still seem to be rather nice.

\section*{Acknowledgements} We would like to thank G. Arutyunov, N. Beisert, and S. Frolov for valuable discussions, and Z. Bajnok, D. Bombardelli and R. Suzuki for valuable comments on the paper. The work by M.L. is supported by the SNSF under project number 200021-137616. The work by S.T. is a part of the ERC Advanced grant research programme No. 246974,  {\it ``Supersymmetry: a window to non-perturbative physics"} and VICI grant 680-47-602.

%%%%%%%%%%%%%%%%%%%%%%%%%%%%%%%%%%%%%%%%%%%%%%%%%%%%%%%%%%%%%%%%%%%%%%%%%%%%%%%%

%%%%%%%%%%%%%%%%%%%%%%%%%%%%%%%%%%%%%%%%%%%%%%%%%%%%%%%%%%%%%%%%%%%%%%%%%%%%%%%%

%%%%%%%%%%%%%%%%%%%%%%%%%%%%%%%%%%%%%%%%%%%%%%%%%%%%%%%%%%%%%%%%%%%%%%%%%%%%%%%%

\appendix

\section{Appendices}

\subsection{Twisted XXX spin chain}
\label{app:XXX}

Let us quickly review the eigenvalues of the twisted transfer matrix for the Heisenberg spin chain with R-matrix
\begin{align}
&r_{12}(u_1,u_2) = \begin{pmatrix}
  1 & 0 & 0 & 0 \\
  0 & b(u_1,u_2) & a(u_1,u_2) & 0 \\
  0 & a(u_1,u_2) & b(u_1,u_2) & 0 \\
  0 & 0 & 0 & 1
\end{pmatrix},
&& a = \frac{U}{u_1-u_2+U},\, b = \frac{u_1-u_2}{u_1-u_2+U}.
\end{align}
It is convenient to write it as
\begin{align}\nonumber
&r_{12}(u_1,u_2)
\,=\,  r_{\alpha\beta}^{\gamma\delta}(u_1,u_2)E^{\alpha}_{\gamma}\otimes E^{\beta}_{\delta}
\,=\,  \frac{u_1-u_2}{u_1-u_2+U}\left[E^\alpha_\alpha\otimes
E^\beta_\beta +\frac{U}{u_1-u_2} E^\alpha_\beta\otimes
E^\beta_\alpha\right]\label{eqn;6vertexComponents},
\end{align}
with $E^\alpha_\beta$ the standard matrix unities. This R-matrix is readily seen to have a $SU(2)$ symmetry. Let us restrict to a twist of diagonal type. Writing the monodromy matrix as
\begin{align}
\mathcal{T} = \prod_{i=1}^{K} r_{0i}(u_0,u_i) = \begin{pmatrix} A(u_0) & B(u_0) \\ C(u_0) & D(u_0)  \end{pmatrix},
\end{align}
then the twisted transfer matrix is of the form
\begin{align}
T = e^{i\gamma}A(u_0) + e^{-i\gamma}D(u_0).
\end{align}
We consider $K$ particles with rapidities $u_i$. The ground state, in this case, is defined as
\begin{eqnarray}
|0\rangle = \bigotimes_{i=1}^{K}\begin{pmatrix}~1 \\ ~0 &
\end{pmatrix}.
\end{eqnarray}
It is easy to check that it is an eigenstate of the twisted transfer matrix. To be a bit more precise, the action of the different elements of the monodromy matrix on $|0\rangle$ is given by
\begin{align}
&A|0\rangle = |0\rangle,&& C|0\rangle = 0,&& D|0\rangle = \prod_{i=1}^{K}b(u_0,u_i)|0\rangle\nonumber.
\end{align}
Thus, $|0\rangle$ is an eigenstate of the transfer matrix with the following eigenvalue
\begin{eqnarray}
e^{i\gamma}+e^{-i\gamma}\prod_{i=1}^{K}b(u_0,u_i).
\end{eqnarray}
The operator $B$ can be considered as a creation operator. It will create all the other eigenstates out of the vacuum. We introduce additional parameters $w_i$ and consider the state
\begin{align}
&|M\rangle := \phi_M(w_1,\ldots,w_M)|0\rangle,&& \phi_M(w_1,\ldots,w_M):=\prod_{i=1}^{M} B(w_i|\vec{u}).
\end{align}
The vacuum corresponds to all spins down and the state $|M\rangle$ corresponds to the eigenstate of the transfer matrix that has $M$ spins turned up.

In order to evaluate the action of the twisted transfer matrix on the state $|M\rangle$, we need the commutation relations between the fields $A,B,D$. From (\ref{eqn;YBE-operators}) we read off that
\begin{eqnarray}
A(u_0)B(w)&=&\frac{1}{b(w,u_0)}B(w)A(u_0)-\frac{a(w,u_0)}{b(w,u_0)}B(u_0)A(w),\nonumber\\
B(w_1)B(w_2)&=&B(w_2)B(w_1),\\
D(u_0)B(w)&=&\frac{1}{b(u_0,w)}B(w)D(u_0)-\frac{a(u_0,w)}{b(u_0,w)}B(u_0)D(w).\nonumber
\end{eqnarray}
From this we can determine exactly when $|M\rangle$ is an
eigenstate of the transfer matrix. By definition we have that
\begin{eqnarray}\label{eqn;XXXinduction}
|M\rangle = B(w_M)|M-1\rangle,
\end{eqnarray}
and this allows us to use induction. By using the identity
\begin{eqnarray}
\frac{1}{b(w_M,u_0)}\frac{a(w_i,u_0)}{b(w_i,u_0)}-\frac{a(w_M,u_0)}{b(w_Mu_0)}\frac{a(w_i,w_M)}{b(w_i,w_M)}
= \frac{a(w_i,u_0)}{b(w_i,u_0)}\frac{1}{b(w_M,w_i)}
\end{eqnarray}
in (\ref{eqn;XXXinduction}) we can prove
\begin{eqnarray}
A(u_0)\phi_M(w_1,\ldots,w_M) &=& \prod_{i=1}^{M}\frac{1}{b(w_i,u_0)}\phi_M(w_1,\ldots,w_M)A(u_0)\\
&&-\sum_{i=1}^{M}\left[\frac{a(w_i,u_0)}{b(w_i,u_0)}\prod_{j=1,j\neq
i}^{M}\frac{1}{b(w_j,w_i)}\hat{\phi}_M A(w_i)\right],\nonumber
\end{eqnarray}
where $\hat{\phi}_M $ stands for $\phi_M(\ldots,w_{i-1},u_0,w_{i+1},\ldots)$. There is a similar relation for the commutator between $D$ and $B$. By using these relations we find that
\begin{eqnarray}
T|M\rangle &=&
\phi_M(w_1,\ldots,w_M)\left\{e^{i\gamma}A(u_0)\prod_{i=1}^{M}\frac{1}{b(w_i,u_0)}
+ e^{-i\gamma} D(u_0)\prod_{i=1}^{M}\frac{1}{b(u_0,w_i)}\right\}|0\rangle \nonumber\\
&&-\sum_{i=1}^{M}\left[\frac{a(w_i,u_0)}{b(w_i,u_0)}\, \hat{\phi}_M \left\{ e^{i\gamma}\prod_{j\neq
i}\frac{1}{b(w_j,w_i)} A(w_i)-e^{-i\gamma}\prod_{j\neq i}\frac{1}{b(w_i,w_j)} D(w_i)\right\}\right]|0\rangle.\nonumber
\end{eqnarray}
From this we find that $|M\rangle$ is an eigenstate of the transfer matrix with eigenvalue
\begin{eqnarray}\label{eqn;eigenvalue6V}
\Lambda^{(6v)}(u_0|\vec{u}) = e^{i\gamma}\prod_{i=1}^M \frac{1}{b(w_i,u_0)}+ e^{-i\gamma} \prod_{i=1}^M \frac{1}{b(u_0,w_i)}\prod_{i=1}^K b(u_0,u_i)
\end{eqnarray}
provided that the auxiliary parameters $w_i$ satisfy the following equations
\begin{eqnarray}\label{eqn;AuxEqns6V}
e^{2i\gamma}\prod_{i=1}^K b(w_j,u_i)=\prod_{i=1,i\neq j}^M \frac{b(w_j,w_i)}{b(w_i,w_j)}.
\end{eqnarray}
This now completely determines the spectrum of the 6-vertex model.

\subsection{Twisted generating functional} \label{app:genfunctional}

As an alternative to our method above, the undeformed transfer matrix can be computed by using a so-called generating functional \cite{Beisert:2006qh,arXiv:0902.0183}. Also, it appears that this method can be extended to twisted transfer matrices. In \cite{Gromov:2010dy} the twist describing $\beta$-deformed theory was implemented on the level of the generating functional for the transfer matrix. They introduced parameters $\tau_i$ in the functional $W$
\begin{align}
W = \frac{1}{1-\tau_1 D \frac{B^{-[-]}Q^{[+]}_1}{B^{+[-]}Q^{[-]}_1} D}\! \left[1-\tau_2 D\frac{Q^{[+]}_1Q_2^{[-2]}}{Q^{[-]}_1Q_2}D\right] \!\!\! \left[1-\tau_3D\frac{Q_2^{[+2]}Q_3^{[-]}}{Q_2Q^{[+]}_3}D\right] \! \frac{1}{1-\tau_4 D\frac{R^{+[+]}Q^{[-]}_3}{R^{-[+]}Q^{[+]}_3}D},
\end{align}
which generates the different transfer matrices by expanding in the shift operator $D=e^{-\frac{i}{g}\partial_u}$ according to
\begin{align}
& W = \sum_{a}D^aT_{a,1}D^a, && W^{-1} = \sum_{s}D^sT_{1,s}D^s.
\end{align}
The various functions in $W$ are defined as
\begin{align}
& R^\pm = \prod_{n=1}^{K_4}  \frac{x(v) - x^{\mp}_{n} }{\sqrt{x^{\mp}_{n} }} ,
&& B^\pm = \prod_{n=1}^{K_4}  \frac{\frac{1}{x(v)} - x^{\mp}_{n} }{\sqrt{x^{\mp}_{n} }} ,
&& Q_j = \prod_{n=1}^{K_j} v-u_{n,j},
\end{align}
and we use the notation $f^{[+a]}(u) = f(u + ia/g)$  ({\it i.e.} $D f^{[a]} = f^{[a-1]}D$). The parameters $x^\pm_n$ correspond to the physical Bethe roots and the function $x(v)$ is related to the parameters of a mirror particle
\begin{align}
x(v+\frac{a i}{g}) + \frac{1}{x(v+\frac{a i}{g})} - x(v-\frac{a i}{g}) + \frac{1}{x(v-\frac{a i}{g})} = \frac{a i }{g}.
\end{align}
Note also that our expression for $W$ is written in terms of the $\alg{psu}(2,2|4)$ roots $K_i$.

The drawback of this approach is that the meaning of the twist parameters in this setting are not clear and need to be fixed a posteriori by knowledge of the twisted Bethe equations\footnote{This is quite the contradistinction to the usual relation between a transfer matrix and the corresponding Bethe equations.}. The authors of \cite{Gromov:2010dy} argued that requiring the Y-functions constructed from these transfer matrices to be real yields that $\tau_2$ and $\tau_3$ as well as $\tau_1$ and $\tau_4$ form complex conjugate pairs. We will see that this is indeed the case.

Let us now compare the transfer matrix obtained by the above described fusion procedure with our explicit derived transfer matrix \eqref{eqn;FullEignvalue}. First we consider the case $T_{1,1}$. Working out the expansion of $W$ to this order is not too complicated and gives
\begin{align}
T_{1,1} = \frac{R^{+[+]}}{R^{-[+]}}\frac{Q_3^{[-]}}{Q_3^{[+]}} \! \left[\tau_4 + \tau_1 \frac{Q_{13}^{[+]}}{Q_{13}^{[-]}} \frac{R^{-[+]}B^{-[-]}}{R^{+[+]}B^{+[-]}}
- \frac{R^{-[+]}}{R^{+[+]}} \left(\! \tau_2 \frac{Q_2^{[-2]}}{Q_2} \frac{Q_{13}^{[+]}}{Q_{13}^{[-]}} + \tau_3 \frac{Q^{[+2]}_2}{Q_2}   \right)\right]\! ,
\end{align}
where $Q_{13} = Q_1 Q_3$. If we take into account the identification between the roots discussed in section \ref{sec:BAE}, we see that this exactly agrees with \eqref{eqn;FullEignvalue} up to overall normalization if we set
\begin{align}
\label{eq:taumatch}
&\tau_1^* = \tau_4 = e^{i\alpha}, && \tau_2^* = \tau_3 = e^{i\beta}.
\end{align}
For the generic $T_{a,1}$ the comparison becomes more involved. However, thanks to the presence of the twist parameters it is easy to compare terms. For instance the term proportional to $e^{i(s-2k-1)\alpha}e^{-i\beta}$ is given by
\begin{align}
\left[D \frac{B^{-[-]}Q^{[+]}_1}{B^{-[+]}Q^{[-]}_1} D\right]^{k}\left[D\frac{Q^{[+]}_1Q_2^{[-2]}}{Q^{[-]}_1Q_2}D\right]\left[ D\frac{R^{+[+]}Q^{[-]}_3}{R^{-[+]}Q^{[+]}_3}D\right]^{s-k-1} \, .
\end{align}
We can then easily compare this against the relevant term from \eqref{eqn;FullEignvalue} after choosing the appropriate overall normalization. Doing this we find perfect agreement after we take into account the relation between $\lambda_{\pm}$ and the shifted parameters as given in \cite{Arutyunov:2009iq}. This proves that he transfer matrix can be equivalently computed by using the generating functional approach.

\subsection{Canonical and Hybrid TBA equations}
\label{app:hybridTBA}

In this appendix we show how the ground state solution is fixed by the canonical TBA equations and how the chemical potentials naturally disappear from the hybrid form of the TBA equations for $Y_+ Y_-$ and $Y_Q$. We will discuss the canonical equations for $w$ strings; for $vw$ strings the story is identical.

\subsubsection*{Canonical equations}

For constant solutions, the ground state canonical TBA equations for $w$ strings in exponential form are given by
\begin{equation}
Y^\circ_{M|w} = e^{-\mu_{M|w}} \left(\prod_{N=1}^{M-1}(1+ \frac{1}{Y^\circ_{N|w}})^{2N} \right) (1+ \frac{1}{Y^\circ_{M|w}})^{2M-1} \left(\prod_{N=M+1}^{\infty}(1+ \frac{1}{Y^\circ_{N|w}})^{2M}\right) \, .
\end{equation}
Next we write $q=e^z$ and write the asymptotic solution (\ref{eq:YwGSasympt}) as
\begin{equation}
Y^\circ_{M|w} = \frac{\sinh{Mz}\sinh{(M+2)z}}{\sinh^2{z}} \, .
\end{equation}
Now we can readily evaluate the following partial sums
\begin{align}
\prod_{N=1}^{M-1}(1+ \frac{1}{Y^\circ_{N|w}})^{2N} & = \frac{\sinh^{2M}{Mz}}{\sinh^2{z}\sinh^{2M-2}{(M+1)z}} \, ,\\
\prod_{N=1}^{k}(1+ \frac{1}{Y^\circ_{N|w}}) & = \frac{\sinh{2 z}}{\sinh{z}} \frac{\sinh{(k+1)z}}{\sinh{(k+2)z}} \, .
\end{align}
Taking the limit $k\rightarrow \infty$ on the second sum for $\mbox{Re}(z)<0$ gives
\begin{equation}
\prod_{N=1}^{\infty}(1+ \frac{1}{Y^\circ_{N|w}})^{2M} = 4^M \cosh^{2M}{z} \, e^{2Mz}\, ,
\end{equation}
and now putting everything together we get
\begin{align}
Y^\circ_{M|w} \, = & \, \, e^{-\mu_{M|w}}  \frac{\sinh^{2M}{Mz}}{\sinh^2{z}\sinh^{2M-2}{(M+1)z}} \left(\frac{\sinh^{2}{(M+1)z}}{\sinh{Mz}\sinh{(M+2)z}}\right)^{2M-1} \nonumber \\
 & \,\, \quad \quad \frac{\sinh^{2M}{z}}{\sinh^{2M}{2 z}} \frac{\sinh^{2M}{(M+2)z}}{\sinh^{2M}{(M+1)z}} 4^M \cosh^{2M}{z} e^{2Mz} \nonumber \\
 = & \,\, 4^M \cosh^{2M}{z} \, \sinh^{2M-2}{z} \, \sinh^{-2M}{2z} \, \sinh{Mz}\, \sinh{(M+2)z} \, \, e^{2Mz-\mu_{M|w}} \nonumber \\
 = & \,\,Y^\circ_{M|w} \, e^{2Mz-\mu_{M|w}} \, ,
\end{align}
showing that since $\mu_{M|w} = 2 i M (\alpha+i \epsilon)$ we should take $z = i (\alpha+i \epsilon)$. For $vw$ strings an identical computation gives $z = i (\beta + i \epsilon)$, and so we obtain the asymptotic solution (\ref{eq:YwGSasympt}-\ref{eq:YQasympt}).

\subsubsection*{Hybrid equations}

In order to derive the Hybrid equations, following \cite{arXiv:1111.5304} we extend the considerations of \cite{AFS09} to the case of non-zero chemical potentials, and indicate the extra terms appearing in the derivation. To derive the hybrid equations we need to compute infinite sums of the form
\begin{equation}
\log \big(1 + \frac{1}{Y_{M|(v)w}}\big) \star \mathcal{K}_M \, ,
\end{equation}
where for our applications $\mathcal{K}_M$ is some kernel which satisfies the schematic identity
\begin{equation}
\label{eq:kernelid}
(K+1)^{-1}_{MN} \star \mathcal{K}_N \equiv \mathcal{K}_M - I_{MN} s \star \mathcal{K}_N = \delta\mathcal{K_M} \, .
\end{equation}
We start by rewriting the simplified equations for $w$ strings
\begin{equation}
\log Y_{M|w} = \, I_{MN}\log(1+Y_{N|w})\star s+
\delta_{M1}\log\frac{1-\frac{1}{Y_{-}}}{1-
\frac{1}{Y_{+}}}\hat{\star}s \,,
\end{equation}
in the form
\begin{align}
\log Y_{M|w} - I_{MN}\log(Y_{N|w}) \star s  =& \, I_{MN}\log\big(1+\frac{1}{Y_{N|w}}\big)\star s+
\delta_{M1}\log\frac{1-\frac{1}{Y_{-}}}{1-\frac{1}{Y_{+}}}\hat{\star}s \,, \\
=& \log Y_{M|w}^{r} - I_{MN}\log(Y^{r}_{N|w}) \star s \, ,
\end{align}
where we have introduced regularized Y-functions $Y_{M|w}^{r} = Y_{M|w} e^{\mu_{M|w}}$ which asymptote to one at large $M$ \textit{cf.} (\ref{eq:YwGSasympt}) given our regulator, without changing the equality. The reason for introducing these regularized Y-functions is that we would like to convolute this equation with $\mathcal{K}_M$ and sum over $M$. For the deformed Y-functions the left hand side of the resulting equation would be the sum of two divergent terms whose sum is regular; the regularized Y-functions simply make this manifest. Using the identity (\ref{eq:kernelid}) we obtain
\begin{equation}
\log Y_{Q|w}^{r} \star \delta \mathcal{K}_Q = \log\big(1+ \frac{1}{Y_{M|w}}\big) \star \mathcal{K}_M - \log \big(1+ \frac{1}{Y_{Q|w}}\big) \star \delta \mathcal{K}_Q +
\log\frac{1-\frac{1}{Y_{-}}}{1-\frac{1}{Y_{+}}}\hat{\star}s \star \mathcal{K}_1 \, .
\end{equation}
Combining the terms appropriately we get
\begin{equation}
\log\big(1+ \frac{1}{Y_{M|w}}\big) \star \mathcal{K}_M  = \log(1+ Y_{Q|w}) \star \delta \mathcal{K}_Q -
\log\frac{1-\frac{1}{Y_{-}}}{1-\frac{1}{Y_{+}}}\hat{\star}s \star \mathcal{K}_1  +  \mu_{Q|w} \star \delta \mathcal{K}_Q\, .
\end{equation}
So we see that the only modification to the identity for the infinite sum over $Y_{M|w}$ functions is the addition of the term $\mu_{Q|w} \star \delta \mathcal{K}_Q$ to the $\log(1+ Y_{Q|w}) \star \delta \mathcal{K}_Q$ contribution. Similarly for $vw$ strings we have to add the term $\mu_{Q|vw} \star \delta \mathcal{K}_Q$ to the $\log(1+ Y_{1|vw}) \star \delta \mathcal{K}_Q$ term.

All that now remains is to identify the kernels $\delta \mathcal{K}$ relevant in the derivation of the hybrid equation for $Y_+ Y_-$ and the one for $Y_Q$, and add the corresponding terms along with the chemical potentials originally in the equation to the undeformed hybrid equations. In the case of $Y_+ Y_-$, $\delta \mathcal{K}_Q = \delta_{Q,1} s$ \textit{cf.} the simplified equation (\ref{eq:sTBAyp}). This means we have to add $2(\mu_{1|vw}-\mu_{1|w})\star s  = \mu_{1|vw}-\mu_{1|w}$ to the chemical potential $-2\mu_y$ of the (doubled) $y$-particles. Now noting that $2\mu_y = \mu_{1|vw}-\mu_{1|w}$ as follows from table \ref{tab:chempot}, we see that the total contribution of the chemical potentials to the hybrid equation for $Y_+ Y_-$ is zero. Next, for $Y_Q$ we have $\delta \mathcal{K}_N = \delta_{N,Q-1} s + \delta_{N,1} s \hat{\star} K_{yQ}$, meaning we add $\mu_{Q-1|vw} \star s + \mu_{1|vw}\star s \hat{\star} K_{yQ} = \tfrac{1}{2} (\mu_{Q-1|vw} + \mu_{1|vw}) = \tfrac{1}{2} \mu_{Q|vw} $ for both the left and right sectors to the chemical potential $-\mu_Q$. This gives a total of $-\mu_Q + \tfrac{1}{2} (\mu^+_{Q|vw} + \mu^-_{Q|vw})$ which is indeed zero for the chemical potentials in table \ref{tab:chempot}.

%%%%%%%%%%%%%%%%%%%%%%%%%%%%%%%%%%%%%%%%%%%%%%%%%%%%%%%%%%%%%%%%%%%%%%%%%%%%%%%%

%%%%%%%%%%%%%%%%%%%%%%%%%%%%%%%%%%%%%%%%%%%%%%%%%%%%%%%%%%%%%%%%%%%%%%%%%%%%%%%%

%%%%%%%%%%%%%%%%%%%%%%%%%%%%%%%%%%%%%%%%%%%%%%%%%%%%%%%%%%%%%%%%%%%%%%%%%%%%%%%%

\end{document}